\documentclass[ApJL,twocolumn]{aastex62}
\usepackage{amssymb, amsmath, graphicx, dcolumn, color, units, xspace, mathtools, physics, tensor, bm, lipsum, revsymb, acronym}
\usepackage[normalem]{ulem}
\usepackage[caption=false]{subfig}
\usepackage{fontawesome5} 

\newcommand{\Msun}{\ensuremath{\,\rm{M}_{\odot}}\xspace}
\newcommand{\AU}{\ensuremath{\,\rm{au}}\xspace}
\newcommand{\Zsun}{\ensuremath{\,\rm{Z}_{\odot}}\xspace}

\newcommand{\Nmodels}{\ensuremath{560}\xspace}

\newcommand{\SFRD}{\ensuremath{\mathcal{S}(Z,z)}\xspace}

\newcommand{\MBHoneZAMS}{\ensuremath{m_{\rm{A}}}\xspace}
\newcommand{\Mchirp}{\ensuremath{\mathcal{M}_{\rm{chirp}}}\xspace}
\newcommand{\q}{\ensuremath{q}\xspace}
\newcommand{\chiBH}{\ensuremath{\chi_{\rm{i}}}\xspace}
\newcommand{\chiBHone}{\ensuremath{\chi_{\rm{1}}}\xspace}
\newcommand{\chiBHtwo}{\ensuremath{\chi_{\rm{2}}}\xspace}

\newcommand{\MBHtwoZAMS}{\ensuremath{m_{\rm{B}}}\xspace}
\newcommand{\pMRR}{\ensuremath{\rm{f}_{\rm{MRR}}}\xspace}

\acrodef{BHNS}{black hole--neutron star}
\acrodef{NSNS}{binary neutron star}
\acrodef{BHBH}{binary black hole}
\acrodef{NS}{neutron star}
\acrodef{BH}{black hole}
\acrodef{BH--NS}{black hole-neutron star}
\acrodef{GW}{gravitational wave}
\acrodef{MRR}{mass ratio reversal}

\definecolor{ForestGreen}{RGB}{34,139,34}
\definecolor{Awesome}{rgb}{1.0, 0.13, 0.32}

\newcommand{\SPA}{School of Physics and Astronomy, Monash University, Clayton VIC 3800, Australia}
\newcommand{\OzGravMonash}{OzGrav: The ARC Centre of Excellence for Gravitational Wave Discovery, Clayton VIC 3800, Australia}
\newcommand{\OzGravSwin}{OzGrav: The ARC Centre of Excellence for Gravitational Wave Discovery, Hawthorn VIC 3122, Australia}
\newcommand{\Swin}{Centre for Astrophysics and Supercomputing, Swinburne University of Technology, Hawthorn VIC 3122, Australia}
\newcommand{\CfA}{Center for Astrophysics \textbar{} Harvard $\&$ Smithsonian,
60 Garden St., Cambridge, MA 02138, USA}

\shorttitle{Mass Ratio Reversals in Gravitational Wave Detections}
\shortauthors{Broekgaarden, Stevenson, and Thrane}

\begin{document}

\title{Signatures of mass ratio reversal in gravitational waves from merging binary black holes}

\author[0000-0002-4421-4962]{Floor S. Broekgaarden}
\affiliation{\CfA}
\author[0000-0002-6100-537X]{Simon Stevenson}
\affiliation{\Swin}
\affiliation{\OzGravSwin}
\author[0000-0002-4418-3895]{Eric Thrane}
\affiliation{\SPA}
\affiliation{\OzGravMonash}
\email{floor.broekgaarden@cfa.harvard.edu, spstevenson@swin.edu.au, \\ eric.thrane@monash.edu}
\date{\today}

\begin{abstract}
The spins of merging binary black holes offer insights into their formation history.
Recently it has been argued that in isolated binary evolution of two massive stars the firstborn black hole is slowly rotating, whilst the progenitor of the second-born black hole can be tidally spun up if the binary is tight enough.
Naively, one might therefore expect that only the less massive black hole in merging binaries exhibits non-negligible spin.
However, if the mass ratio of the binary is ``reversed'' (typically during the first mass transfer episode), it is possible for the tidally spun up second-born to become the more massive black hole.
We study the properties of such mass-ratio reversed (MRR) binary black hole mergers using a large set of 560 population synthesis models. %
We find that the more massive black hole is formed second in $\gtrsim 70\%$ of binary black holes observable by LIGO, Virgo, and KAGRA for most model variations we consider, with typical total masses $\gtrsim 20$\,M$_\odot$ and mass ratios  $q = m_2 / m_1 \sim 0.7$ (where $m_1 > m_2$).
The formation history of these systems typically involves only stable mass transfer episodes.
The second-born black hole has non-negligible spin ($\chi > 0.05$) in up to $25\%$ of binary black holes, with among those the more (less) massive black hole  spinning in $0\%$--$80\%$ ($20\%$--$100\%$) of cases, varying greatly in our models.
We discuss our models in the context of several observed gravitational-wave events and the observed mass ratio - effective spin correlation.\\
\end{abstract}

\section{Introduction}
\label{sec:intro}
The population of binary black hole mergers observed by Advanced LIGO \citep{TheLIGOScientificDetector:2014jea}, Virgo \citep{TheVirgoDetector:2014hva} and KAGRA \citep[][]{KAGRA:2018plz} is rapidly increasing. 
The inferred merger rates of binary black holes, and their observed mass and spin distributions offer insights into their formation history \citep[e.g.,][]{Stevenson:2015bqa,Stevenson:2017dlk,Rodriguez:2016vmx,Talbot:2017yur,Vitale:2015tea,Farr:2017uvj}.
By comparing the data with theoretically synthesized populations, we can learn about their massive-star progenitors.

The birth spins of black holes are uncertain. 
Recently, several authors have argued that angular momentum transport within massive stars is efficient \citep[][]{Spruit:2001tz,Fuller:2019MNRAS}, as also supported by observations from astroseismology \citep{2014MNRAS.444..102K,2014A&A...564A..27D,2018A&A...616A..24G} and gravitational waves \citep{Belczynski:2017gds, Zevin:2020gxf}.   Under this assumption, and the assumption of Eddington-limited accretion, we expect the firstborn black hole to be almost non-rotating \citep[with dimensionless spin $\chi \lesssim 0.01$;][]{Qin:2018vaa,Fuller:2019sxi} as the majority of the progenitors' angular momentum will have been transported from its core to its envelope, and subsequently removed through mass transfer and stellar winds.

The same may not be true for the second-born black hole, however. 
Immediately prior to the formation of the second black hole, the binary consists of a black hole and a helium (Wolf-Rayet) star with a relatively short orbital period.
Tides exerted on the helium star by the black hole can synchronise the rotation of the helium star with the orbital period, leading to a rapidly rotating helium star that may subsequently collapse to form a rapidly rotating black hole \citep{Kushnir:2016zee,Hotokezaka:2017ApJ,2018MNRAS.473.4174Z,Qin:2018vaa,Bavera:2019,Belczynski:2017gds}.
 
Naively, if only the second-born black hole can be rapidly rotating, one might expect the more massive black hole (with mass $m_1$) to be formed first with no spin ($\chi_1 \sim 0$)\footnote{Where we use the notation $\chi_1$ ($\chi_2$) for the spin of the more (less) massive binary black hole. See \citet{Biscoveanu:2020are} for an alternative parameterization.}.
This intuition originates from a naive picture of massive binary evolution in which two massive stars never interact: first, the more massive star forms the more massive black hole, since the relation between the initial mass of a star and the black hole mass is believed to be monotonic over a wide range of initial masses \citep[e.g.,][]{Belczynski:2010ApJ,COMPAS:2021methodsPaper}.
Second, the more massive black hole forms first, since the initially more massive star has the shorter lifetime \citep[e.g.,][]{Agrawal:2020MNRAS,COMPAS:2021methodsPaper}.

However, binary evolution complicates this picture. 
Mass transfer from the initially more massive star\footnote{Stellar theorists typically refer to the initially more massive star as the ``primary'' and the less massive star as the ``secondary''. We avoid this jargon since gravitational-wave astronomers refer to the more massive black hole as the ``primary.'' This leads to confusion when the initially more massive star produces the less massive black hole.} to its companion can lead to the mass ratio of the binary being reversed, potentially leading to the initially lower mass star becoming more massive than the primary ever was. 
For binaries that go on to form merging binary black holes, this can ultimately result in the second black hole to form in the binary being more massive than the first \citep[e.g.,][]{Gerosa:2013laa,Stevenson:2017tfq,Zevin:2022wrw}\footnote{Algol binaries are a classic example of binaries that have undergone mass ratio reversal. Mass ratio reversal has also been suggested for  binary neutron stars \citep[e.g.,][]{1996A&A...309..179P}, neutron star--black hole binaries \citep[e.g.,][]{2004MNRAS.354L..49S}, and neutron star--white dwarf binaries  \citep[e.g.,][]{2000A&A...355..236T,2018A&A...619A..53T} with observational support from young neutron stars in neutron star--white dwarf binaries \citep[e.g.,][]{2000ApJ...543..321K, 2000A&A...355..236T,2018MNRAS.476.4315N,Venkatraman:2020}.}. 
This mass ratio reversal can have important implications for the spins of the black holes in the binary.  
If the binary has undergone mass ratio reversal, this leads to the intriguing possibility that the (progenitor of the) more massive black hole is the one that may be tidally spun up, leading to $\chi_1 \geq 0$.

In this paper, we investigate the frequency and properties of \ac{MRR} systems in the population of merging binary black holes formed through isolated binary evolution with a goal of making testable predictions for gravitational-wave astronomers. We use a large set of models to test how robust our predictions are to uncertainties in both our treatment of massive binary evolution, and of the cosmic star formation history.
We begin by introducing our methodology in Section~\ref{sec:binpop}, where we also describe the typical formation history of a merging binary black hole that has undergone \ac{MRR}.
In Section~\ref{sec:resI}, we show that the majority of observable binary black holes undergo MRR in most of our models.
We describe the properties of the \ac{MRR} versus non-\ac{MRR} systems.
In Section~\ref{sec:resII:spins}, we illustrate that, within the subset of binary black hole mergers with non-negligible spin, \ac{MRR} (non-\ac{MRR}) systems are identifiable by having non-negligible $\chi_1$ ($\chi_2$).
We describe the chirp mass and mass ratio properties of these sub-populations. 
We end with a discussion in Section~\ref{sec:discussion} and 
conclude in Section~\ref{sec:conclusions}.

\section{Method}
\label{sec:binpop}
\subsection{Population Synthesis Simulations}

We study merging binary black holes formed from the evolution of isolated massive binary stars (the `isolated binary evolution' channel) using the publicly available simulations presented in \citet{Broekgaarden:2021iew,Broekgaarden:2021efa}. 
These simulations were performed using the rapid binary population synthesis suite COMPAS\footnote{Compact Object Mergers: Population Astrophysics and Statistics -- \url{https://compas.science}} \citep{Stevenson:2017tfq,Vigna-Gomez:2018dza,Broekgaarden:2019qnw,COMPAS:2021methodsPaper}, which
 employs simple parameterized models of single \citep{Hurley:2000MNRASSSE} and binary stellar evolution \citep{Hurley:2002MNRASBSE} in order to rapidly evolve large populations of binaries.

The simulations from \citet{Broekgaarden:2021iew,Broekgaarden:2021efa} present 560 model realizations, exploring a large range of  uncertainties underlying population synthesis models. This includes 20 variations in assumptions related to uncertain stages of massive (binary) star evolution such as mass transfer, supernovae kicks, common envelope evolution and Wolf-Rayet winds (see our online \href{https://github.com/FloorBroekgaarden/MRR_Project/blob/main/otherFiles/DCO_table_detailed.png}{Table}).
Each stellar evolution variation is combined with 28 different  assumptions about the metallicity-dependent star formation rate density \SFRD, which is a function of metallicity ($Z$) and redshift ($z$) and describes the amount of star formation and the distribution of the birth metallicities as a function of cosmic time. Both the stellar evolution and  \SFRD uncertainties have been shown by a number of authors to be an important ingredient in predicting populations of merging binary black holes \citep[e.g.][]{Neijssel:2019,Tang:2019qhn,Santoliquido:2020bry,Briel:2021bpb,Broekgaarden:2021efa}.  

To make a meaningful comparison with gravitational-wave data we convert the COMPAS population into an \textit{astrophysical} population of all local (redshift $z \approx 0$) sources and a \textit{detectable} population. 
We do this using the weights based on Equation 2 and 5 from \citet{Broekgaarden:2021efa}.
For simplicity, we assume for the detectable population a detector network consisting of LIGO, Virgo and KAGRA at design sensitivity (from hereon `LVK'), which can see more distant merger events than the current network.
For the remainder of this paper we refer to these weighted rates when we discuss the astrophysical or detectable population.  We refer the readers to \citet{Broekgaarden:2021iew,Broekgaarden:2021efa} for more details. 

\subsection{Selecting Mass Ratio Reversals}
\label{method:mass_ratio_reversal}
From the population of binary black hole mergers we obtain the subset of \acp{MRR} by selecting the systems for which the black hole that formed second from the initially less massive star (B)  is more massive than the black hole formed first from the initially more massive star (A). 
In other words, we define \ac{MRR} as systems that satisfy the condition $\MBHtwoZAMS > \MBHoneZAMS$ at the moment of the binary black hole formation such that the binary system has `reversed' its mass ratio between its formation as a binary system and its merger as a binary black hole\footnote{There are a negligible number of systems in which significant mass transfer causes \MBHtwoZAMS to undergo supernova first, in which case we define it to be MRR if $\MBHtwoZAMS  < \MBHoneZAMS$, as we are interested in whether the most massive black hole formed second.}. 

\subsection{Modelling Black Hole Spin}
\label{method:black_hole_spin_modelling}
We model the formation of black hole spin following the methodology of \citet{Bavera:2019,Bavera:2020uch}. 
We assume that the firstborn black hole in the binary system always has zero spin, corresponding to efficient angular momentum transport in the star \citep[][]{Qin:2018vaa}. 
For the second-born black hole, we assume that it can be formed spinning if its progenitor helium star has been tidally spun up. 
This requires the binary to go through a black hole--Wolf Rayet phase with an orbital period of less than 1 day. 
For these systems we use the prescription for the second-born black-hole spin given by Equation 1 and 2 in \citet{Bavera:2021evk}, which they based on detailed MESA simulations. This prescription assigns a black hole spin based on the orbital separation and the Wolf-Rayet stellar mass. 

To implement this approximation within our simulations, we use the coefficients given for Wolf-Rayet stars at helium depletion. In addition, a few of our binary systems have properties at helium depletion that fall outside of the range of Wolf-Rayet star period and mass given in \citet{Bavera:2021evk}. In these cases we assign the spin value using the most similar binary within the allowed range.  
We further note that although the \citet{Bavera:2021evk} approximation is built for the delayed supernova model prescription from \citet{Fryer:2012ApJ}, we apply it, e.g., to our model (L) that uses the rapid supernova remnant mass prescription. 
We do not expect this to drastically change our results.

\begin{figure}
    \centering
\includegraphics[width=1\columnwidth]{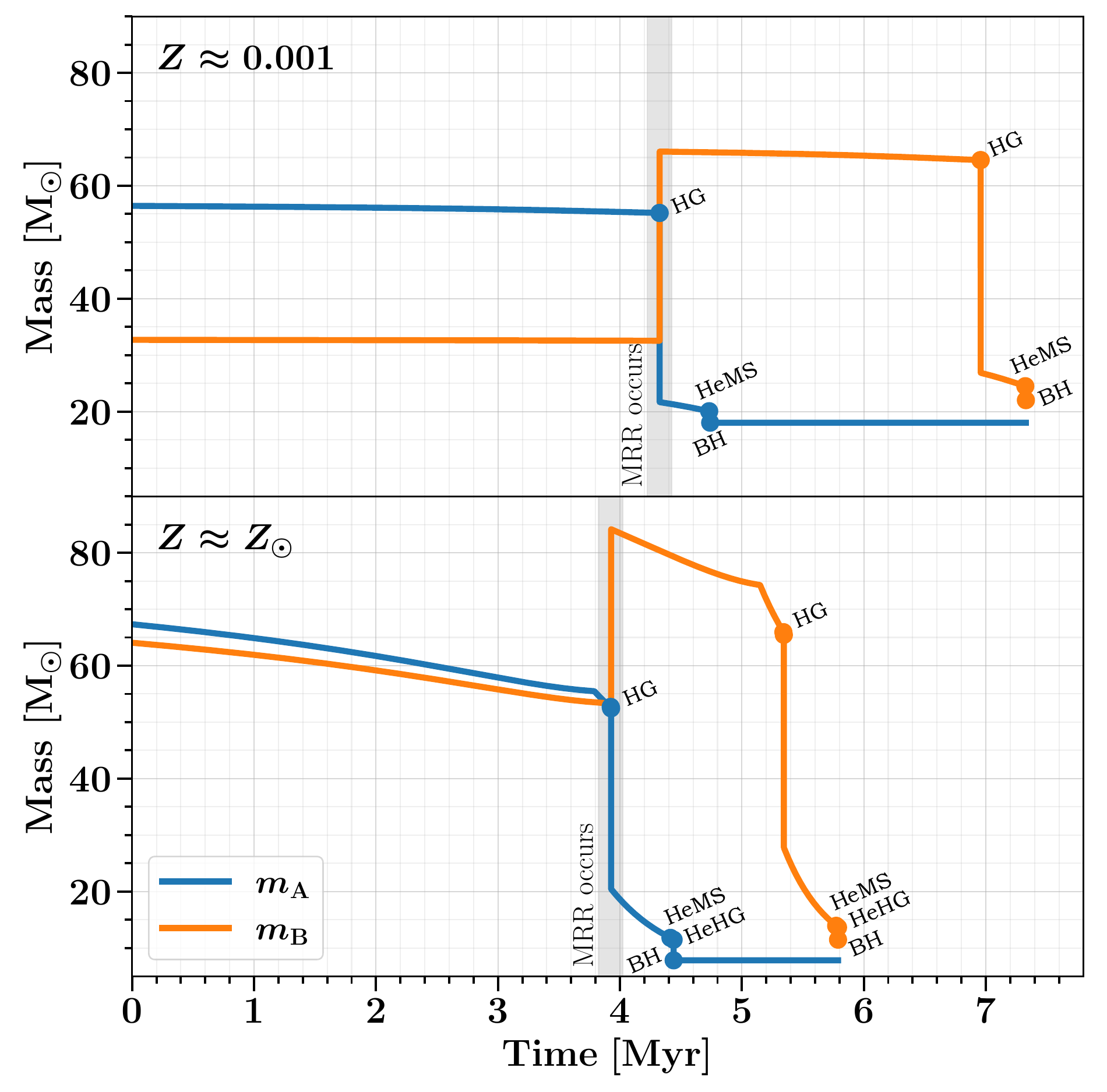}
\caption{Examples of a binary system in our fiducial model realization that undergoes mass ratio reversal (MRR) evolved at low metallicity ($Z\approx0.001$; top panel) and solar metallicity ($Z\approx0.0142$; bottom panel).  Each panel shows the evolution of the initially more massive ($\MBHoneZAMS$) and less massive ($\MBHtwoZAMS$) star as a function of time.  %
Both systems undergo \ac{MRR} during the first mass transfer episode after approximately 4 Myr into their evolution (as indicated by the gray vertical bar).
Transitions to different evolutionary stages are labelled using the acronyms: HG for Hertzsprung gap star, HeMS for helium main sequence star, HeHG for helium Hertzsprung gap star, and BH for black hole. 
More examples are provided on  \href{https://github.com/FloorBroekgaarden/MRR_Project/tree/main/Figure1_individual_MRR_systems_detailed_evolution/detailedPlots/extraPlots}{GitHub}. 
\href{https://github.com/FloorBroekgaarden/MRR_Project/blob/main/Figure1_individual_MRR_systems_detailed_evolution/detailedPlots/Figure_1_MRR.pdf}{\faFileImage} \href{https://github.com/FloorBroekgaarden/MRR_Project/blob/main/Figure1_individual_MRR_systems_detailed_evolution/Individual_System_Evolution_and_MRR_Statistics.ipynb}{\faBook} 
}
\label{fig:DetailedEvolutionMRR}
\end{figure}

\subsection{Formation channels}
\label{sec:formation-channels-example}
We begin by outlining the formation history of a typical progenitor of a merging binary black hole that is a \ac{MRR} system. 
An example is shown in Fig.~\ref{fig:DetailedEvolutionMRR} for a binary formed at low ($Z\approx 0.001$) and solar-like ($Z \approx \Zsun$) metallicity.
Two massive stars are born in a wide binary. 
The examples in Fig.~\ref{fig:DetailedEvolutionMRR} start with a binary at $Z\approx 0.001$  ($Z \approx \Zsun$) metallicity with masses $\MBHoneZAMS \approx 56\Msun$ ($\MBHoneZAMS \approx 67\Msun$),  $\MBHtwoZAMS \approx 33\Msun$ ($\MBHtwoZAMS  \approx 64\Msun$) and a separation of $70 \AU$ ($1000\AU$). 
The initially more massive star evolves off of the main sequence, and fills its Roche lobe whilst crossing the Hertzsprung gap (HG), which occurs in the example systems after approximately 4 Myr.

If the mass ratio of the binary is not too far from unity (e.g., $\MBHoneZAMS / \MBHtwoZAMS  \lesssim 4$ \citealt{Claeys:2014,COMPAS:2021methodsPaper}),  we find that the mass transfer episode is stable \citep[e.g.,][]{Hurley:2002MNRASBSE}, and the envelope of the initially more massive star is removed, with a large fraction being accreted onto the initially more massive star \citep{Hurley:2002MNRASBSE,Schneider:2015ApJ,COMPAS:2021methodsPaper}. 
This is the evolutionary stage in our COMPAS simulations where \ac{MRR} typically occurs. 
After the stable mass transfer phase, the firstborn star has evolved into a stripped helium star (HeMS) of $\MBHoneZAMS \approx 18\Msun$ ($\MBHoneZAMS \approx 11\Msun$) and the second-born star has evolved to a $\MBHtwoZAMS \approx 66\Msun$ ($\MBHtwoZAMS \approx 79\Msun$) for low (solar) metallicity.  
The stripped helium core subsequently collapses to form a black hole of $\MBHoneZAMS \approx 18\Msun$ ($\MBHoneZAMS \approx 8\Msun$).  

Following the formation of the first black hole, the initially less massive star eventually reaches the end of its main sequence and subsequently fills its Roche lobe typically when crossing the Hertzsprung gap (HG).
Once again, depending on various properties such as the evolutionary stage of the donor star and the binary mass ratio, this phase of mass transfer can either be stable, or unstable resulting in common envelope evolution \citep[][]{vandenHeuvel:2017pwp,Neijssel:2019,Bavera:2020uch,Gallegos-Garcia:2021hti,vanSon:2021zpk}. 
In the binary portrayed in Fig.~\ref{fig:DetailedEvolutionMRR}, which is typical of our population synthesis models, we find that this episode of mass transfer is typically stable, as we discuss in more detail in the next section. 

The stable mass transfer onto the black hole companion is highly non-conservative (we assume that the maximum mass accretion rate for a compact object is set by the Eddington limit), leading to drastic shrinking of the binary's orbit. 
Eventually the initially less massive also forms a black hole---in this case of $\MBHtwoZAMS \approx 22\Msun$ ($\MBHtwoZAMS \approx 11\Msun$) for low (solar) metallicity, forming the most massive black hole in the binary black hole system. 
The subsequent \ac{MRR} binary black hole system merges in a Hubble time as a potential source for LVK.
The main difference at low versus solar metallicity is that, at solar metallicity, the stars lose more mass through line driven stellar winds, leading to smaller black hole masses. 

We classify our binary black hole systems into three different categories based on their formation channel (based on, e.g., \citealt{Neijssel:2019,Bavera:2020uch,Broekgaarden:2021iew}).
Namely: if the binary experiences only stable mass transfer we classify the channel as \textbf{ `Only stable mass transfer'}. 
Both binaries in Fig.~\ref{fig:DetailedEvolutionMRR} are examples of this channel. 
On the other hand, if the second, reversed, mass transfer episode is unstable, leading to a common-envelope phase, we classify the channel as \textbf{`Classic Common Envelope'} channel. 
We additionally assume for both channels that the first mass transfer phase must occur when the initially more massive star has evolved off the main sequence to exclude so-called case A mass transfer systems that are typically more poorly modelled in population synthesis simulations. 
We classify all other systems under \textbf{``Other''}.

\section{Results: Mass Ratio Reversal Rates and Properties}
\label{sec:resI}

In this section we study the rate (\S\ref{subsec:pMRR}) and properties  (\S\ref{subsubsec:MRR-properties-DCO-formation}) of \ac{MRR} systems among the population of binary black hole mergers.  

\begin{figure*}
    \centering
\includegraphics[width=1\textwidth]{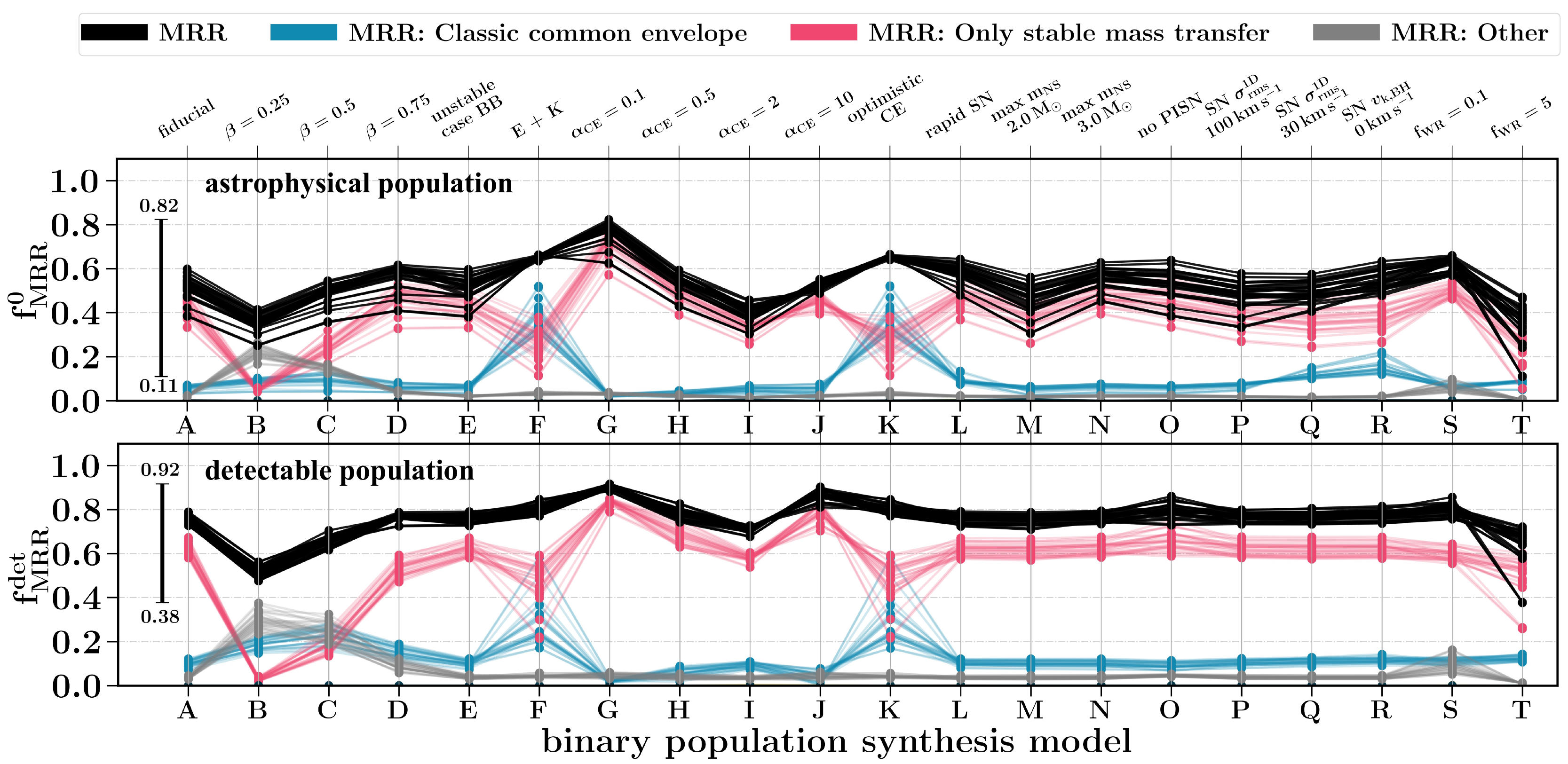}
\caption{Fraction of mass ratio reversals (\acp{MRR}) among all binary black hole mergers (black) expected in the astrophysical (redshift 0; top panel) and gravitational-wave detectable (lower panel) populations for our \Nmodels model realizations.  The fraction of \ac{MRR} binary black holes that go through the classic common envelope (blue), only stable mass transfer (magenta) and other channel (gray) are shown (see \S\ref{sec:formation-channels-example}). We connect models using the same star formation rate density \SFRD with a line for visual purposes only.
   \href{https://github.com/FloorBroekgaarden/MRR_Project/blob/main/Figure2_MRR_Rates_For_All_Models/RelativeMRRrates_Combined.pdf}{\faFileImage} \href{https://github.com/FloorBroekgaarden/MRR_Project/blob/main/Figure2_MRR_Rates_For_All_Models/MRR_Rates_Intrinsic_and_Detectable_summary.ipynb}{\faBook}
  }
 \label{fig:FractionMRRbinaryBHs}
\end{figure*}

\subsection{Fraction of mass ratio reversals among merging binary black holes }
\label{subsec:pMRR}

In Fig.~\ref{fig:FractionMRRbinaryBHs} we show the fraction of binary black holes in each model realization that have undergone mass ratio reversal, \pMRR{}, for the \emph{astrophysical} (top panel) and \emph{detectable} (bottom panel) population. 
The figure also shows the contributions from the different formation channels as defined in \S\ref{sec:formation-channels-example}.  
We find several interesting features in the \ac{MRR} rate results. 

First, we find that for the gravitational-wave detectable population, the fraction of \acp{MRR} lies between $\pMRR^{\rm{det}}$~$\approx0.38$--$0.92$ (panel \ref{fig:FractionMRRbinaryBHs}), indicating that mass ratio reversal is expected to be a common occurrence in LVK's observations.  
Moreover, when excluding a subset of the models with extreme Wolf-Rayet wind factors of $f_{\rm{WR}}=5$ (model T), we find that the fraction of \acp{MRR} is always larger than $\gtrsim 0.5$, with most models predicting a contribution $\pMRR^{\rm{det}} \gtrsim 0.7$. 
In other words, for most models, we expect the more massive black hole formed second for the majority of LVK binary black holes.
We discuss the consequences for LVK measurements of black-hole spin in \S\ref{sec:resII:spins}. 

Second, we find that for the astrophysical binary black hole population, the fraction of \acp{MRR} is lower compared to the detectable population, lying between $\pMRR^{\rm{0}}$~$\approx0.11$--$0.82$ (panel \ref{fig:FractionMRRbinaryBHs}) in all of our models, and $\pMRR^{\rm{0}} \gtrsim 0.3$ when excluding a subset of the models T. 
This fraction is lower compared to the detectable population because, as we will show in \S~\ref{subsubsec:MRR-properties-DCO-formation}, the binaries that undergo \ac{MRR} form typically more massive binary black holes compared to the non-\ac{MRR} systems. 
These more massive MRR systems are easier to  detect with the LVK network, boosting their contribution to the detectable population.

Third, we find that the majority of \ac{MRR} binary black holes form through the only stable mass transfer sub-channel in most of our models.
The exceptions are the models with a mass transfer efficiency parameter $\beta = 0.25$ (models B) and the stellar evolution models with the `optimistic common envelope' assumption (models F and K). 
In the $\beta = 0.25$ models the MRR fraction is drastically suppressed as the first mass transfer phase is relatively non-conservative, which results in limited mass gain by the initially less massive star.
The optimistic common envelope models allow Hertzsprung-Gap star donors that engage in a common-envelope phase to survive this evolutionary stage (in the default `pessimistic common envelope' assumption we assume these systems merge during this phase).
This increases the number of binary black holes, and the (relative) number of \ac{MRR} forming through the classic common envelope channel.

Finally, we find that the fraction of \ac{MRR} binary black holes increases for increasing mass transfer efficiency $\beta$ (models B, C, and D), decreases for increasing common-envelope efficiencies after $\alpha = 0.1$ (models G, H, I) but increases again for $\alpha = 10$ (models J), and is not significantly impacted by our variations in supernovae physics (models L--R).
We discuss in more detail the outlier models in Appendix~\ref{app-fraction-MRR-outliers}.

\subsection{Observable properties of MRR binaries}
\label{subsubsec:MRR-properties-DCO-formation}
\begin{figure*}
    \centering
    \includegraphics[width=1.0\textwidth]{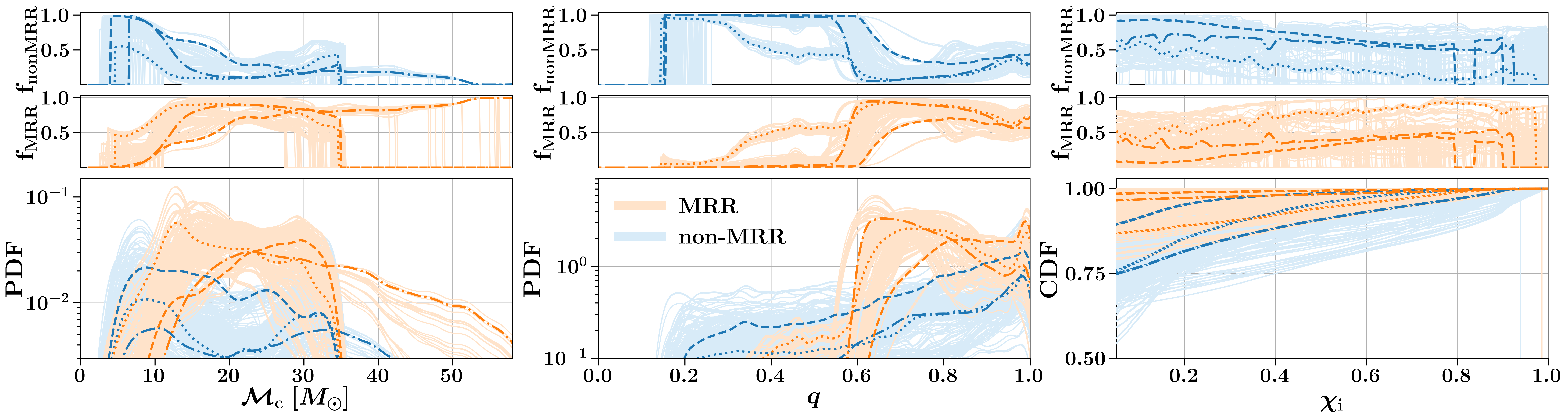}
    \caption{Chirp mass (left panels), mass ratio (middle panels), and dimensionless black hole spin (right panels) for the populations of mass ratio reversed (MRR; orange) and non-MRR (blue) detectable binary black holes for all \Nmodels model realizations. 
    Three realizations, `K123', `O312' and `T231' (see \href{https://github.com/FloorBroekgaarden/MRR_Project/blob/main/README.md}{Github}) are highlighted with a dotted, dash-dotted and dotted curve, respectively.
    The bottom panels show the probability distribution functions (PDFs), except for the black hole spin where we show the cumulative distribution functions (CDFs) for visualization purposes. 
    The two top rows show the fraction of expected detectable \ac{MRR} ($\rm{f}_{\rm{MRR}}$) and non-MRR ($\rm{f}_{\rm{nonMRR}}$) binaries.
    The PDFs are normalized such that for each model realization the MRR and non-MRR PDF sum up to one to show the relative normalization.
   \href{https://github.com/FloorBroekgaarden/MRR_Project/blob/main/Figure3_and_Figure5_BBH_properties_MRR_and_nonMRR_Distribution_plots/super_MRR_split_panel.pdf}{\faFileImage} \href{https://github.com/FloorBroekgaarden/MRR_Project/blob/main/Figure3_and_Figure5_BBH_properties_MRR_and_nonMRR_Distribution_plots/Make_Figure_3_and_5_Distribution_Plots.ipynb}{\faBook}
    }
\label{fig:MRR-vs-not-MRR-BBH-properties}
\end{figure*}

In the previous section we found that \ac{MRR} systems can significantly contribute to the detectable binary black hole population. It is, therefore, interesting to consider whether their properties (i.e., chirp mass, mass ratio, spins) are distinguishable from other binary black hole mergers detected by the LVK. In this section, we compare and contrast these properties in the \ac{MRR} and non-\ac{MRR} binary black hole populations.

First, we find that the detectable \ac{MRR} binary black hole mergers are biased to larger chirp masses, \Mchirp  (and total mass and individual black hole masses) in all our simulations. 
As shown in Fig.~\ref{fig:MRR-vs-not-MRR-BBH-properties}, we see that \ac{MRR} binary black hole systems typically have chirp mass distributions peaking between $\Mchirp \approx 10$--$35\Msun$, whereas the non-\ac{MRR} population peaks around $\Mchirp \approx 5$--$20\Msun$. 
This is also clear from the top panel, which shows that typically between $50$--$100\%$ of the detected binary black hole mergers with chirp mass $\Mchirp \gtrsim 10\Msun$ are expected to be MRRs with a peak around 20\Msun where we expect the large majority to be MRRs. 
This is because MRR binaries are predominantly formed through the stable mass transfer channel (cf. Fig.~\ref{fig:FractionMRRbinaryBHs}), whereas non-MRR binaries are typically formed through the classic common envelope channel. 
Binary black holes formed through stable mass transfer are more massive than those formed through common envelope evolution \citep[][]{Neijssel:2019,vanSon:2021zpk}, leading to this distinction. 
In the chirp mass plot, one group of models stand out by having a long tail of $\Mchirp \gtrsim 35$.
These are our models that do not include pair-instability supernovae (models O), allowing for the formation of black holes with $m_{\rm{BH}}\gtrsim 40\Msun$. 
The MRR binaries dominate this tail. 

Second, MRR binary black holes also have distinct mass ratios ($q$) compared to the non-MRR systems, as shown in Fig.~\ref{fig:MRR-vs-not-MRR-BBH-properties}. 
The MRR systems peak, and typically dominate observations, for $q\gtrsim 0.6$, whereas only the non-MRR can typically form binary black holes with $q \lesssim 0.6$. This is because for MRR binaries more extreme mass ratios require more extreme mass accretion by the initially less massive star, which is typically challenging and is also disfavored to lead to the successful formation of a binary black hole merger (e.g., it requires more extreme mass ratios in earlier phases of the binary, that more likely leads to unstable mass transfer phases making the stars merge). 
More than 50\% of binary black holes with $q \sim 0.7$ are MRR in almost all of our model realizations, whereas for mass ratios $q \sim 0.4$ most models (except the optimistic common envelope models) predict that less than 10\% of binary black holes are MRR.
This preference for mass ratios close to $0.7$ in the MRR population comes from the fact that binaries that undergo MRR are born with mass ratios close to unity (see \S~\ref{subsec:MRR-properties-birth}). 

Let us consider a simplified picture \citep[see discussion in][]{vanSon:2021zpk} to explore this, inspired by Fig.~\ref{fig:DetailedEvolutionMRR}.
Consider a binary born with approximately equal mass components $m_\mathrm{A}^\mathrm{ZAMS} \approx m_\mathrm{B}^\mathrm{ZAMS} = m^\mathrm{ZAMS}$, and let us neglect stellar winds for this simple estimate.
Now assume that all of the envelope (which is typically around 50\% of the stellar mass, $m^\mathrm{ZAMS} / 2$) of the initially more massive star is accreted on to the initially less massive star (i.e., assuming fully conservative mass transfer).
At this stage, the mass ratio of the binary would be around $m_\mathrm{B} / m_\mathrm{A} \sim (3 m^\mathrm{ZAMS} / 2) / (m^\mathrm{ZAMS} / 2) \sim 3$. 
Again neglecting further mass-loss through winds, we take the final mass of the firstborn black hole to be $m_\mathrm{A}^\mathrm{ZAMS} / 2$. 
The envelope of the initially less massive star is then removed through non-conservative stable mass transfer, leaving it with a final mass of $3 m^\mathrm{ZAMS} / 4$.
Hence, the final mass ratio of the binary is around $(3 m^\mathrm{ZAMS} / 2) / (m^\mathrm{ZAMS} / 2) = 3/2$, or equivalently $q \sim 0.7$.

In general (both MRR and non-MRR), binary black holes formed in our models show a preference for close-to-equal masses, with highly asymmetric mass ratios $q < 0.1$ being so rare as to be essentially non-existent \citep{Broekgaarden:2021efa}, cf.  \citet{Olejak:2020oel}. There are two sets of models with contributions from the MRR channel to   extreme mass ratios around $q\approx 0.5$,  which are the models that assume the optimistic common envelope prescription (models F and K).

\begin{figure*}
    \centering
    \includegraphics[width=1\textwidth]{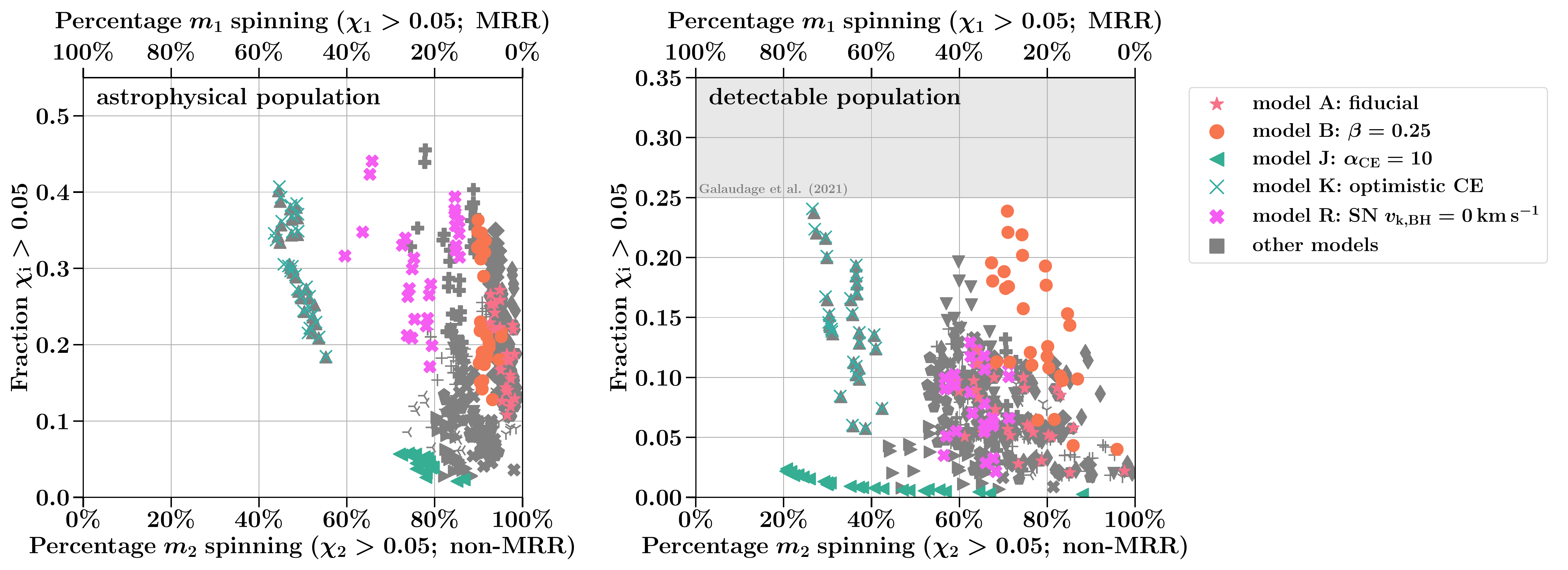}
    \caption{Fraction of the astrophysical (left panel) and detectable (right panel) binary black hole merger population where one of the black holes has a non-negligible spin magnitude ($\chiBH > 0.05$) plotted against  
    the percentage of systems in which the more massive ($m_1$) or less massive ($m_2$) black hole is spinning.
    Each marker style corresponds to one of our 20 stellar evolution variations for the combination with 28 cosmic star formation history models (see \href{https://github.com/FloorBroekgaarden/MRR_Project/blob/main/README.md}{Github}). 
    Five stellar evolution variations are highlighted. 
    Part of the $90\%$ credible interval range of $0.25$--$0.71$ for the fraction of binary black hole mergers with non-negligible spin from \citet{Galaudage:2021rkt} is shown with a gray bar. 
    \href{https://github.com/FloorBroekgaarden/MRR_Project/blob/main/Figure4_spinning_BBH_Rates/Rate_MRR_spins_panels_spin_threshold_0.05.pdf}{\faFileImage} \href{https://github.com/FloorBroekgaarden/MRR_Project/blob/main/Figure4_spinning_BBH_Rates/Plot_fraction_spinning_BBHs.ipynb}{\faBook} 
    }
    \label{fig:frac_spinning}
\end{figure*}

Third, we find in Fig.~\ref{fig:MRR-vs-not-MRR-BBH-properties} that, for both the MRR and non-MRR populations, a significant fraction of the binary black holes contain a black hole with non-negligible spin ($\chi_{\rm{i}} \gtrsim 0.05$).
In our models, it is always the second-born black hole that is spinning, as the spin arises from tidal spin up.
Hence, for MRR binaries, the more massive black hole is spinning ($\chi_{1} > 0$), whereas for non-MRR binaries, it is the less massive black hole that is spinning ($\chi_{2} > 0$).
The fraction of binary black holes with no spinning black hole is typically smaller for the non-MRR population ($\gtrsim 60\%$) compared to the MRR population ($\gtrsim 80\%$). 
This is because non MRR binary black holes typically go through a common-envelope phase, which shrinks the binary to smaller orbital separations compared to a second stable mass transfer phase.
These shorter separations are more likely to allow tidal spin up of the second-born black hole. 
We discuss the implications of mass ratio reversal for black hole spins further in the following section.

\section{Results: Mass Ratio Reversal and Black Hole Spin}
\label{sec:resII:spins}

The population of detectable binary black hole binaries with a non-negligible black hole spin component (identified in  Fig.~\ref{fig:MRR-vs-not-MRR-BBH-properties}) allows for identification of MRR versus non-MRR systems from observations.
Namely, for MRR binary black holes, the more massive black hole has non-negligible spin $\chiBHone > 0.05$\footnote{Where we define in our study `non-negligble spin' as systems with $\chi_{\rm{i}} > 0.05$ to exclude systems with extremely small spin. Changing this to, e.g., 0.01, did not significantly impact our results.}, and the less massive black hole is non-spinning, $\chiBHtwo = 0$, whilst for non-MRR binaries  $\chiBHtwo > 0.05$ and $\chiBHone = 0$, as a result of our assumption that only the second-born black hole can spin up through tides. 
We discuss the implications of this below.

\subsection{Fraction of binary black holes with non-negligible spin }

Fig.~\ref{fig:frac_spinning} shows the fraction of binary black holes with non-negligible spin ($\chiBH > 0.05$), as well as the fraction among those where the non-negligible spin is assigned to the more massive black hole  ($\chiBHone > 0.05$; MRR) or less massive black hole ($\chiBHtwo > 0.05$; non-MRR).  We find the following.

First, for the astrophysical population we find that the fraction of binary black holes with at least one black hole spinning lies between $0$--$0.5$, whereas for the detectable population this becomes $0$--$0.25$. In other words, we expect that at most $25\%$ of the binary black holes detected by LVK will contain a black hole that has non-negligible spin, whereas for the astrophysical population this can reach as high as $50\%$. 
The fraction is higher for the astrophysical rate because this population contains a higher fraction of non-MRR systems (Fig.~\ref{fig:FractionMRRbinaryBHs}), which have a higher contribution of spinning black holes compared to \ac{MRR} systems (Fig.~\ref{fig:MRR-vs-not-MRR-BBH-properties}). Recently, \citet{Galaudage:2021rkt}\footnote{We use the updated fractions quoted in their \href{https://arxiv.org/pdf/2109.02424.pdf}{erratum}} inferred that the fraction of observed binary black holes with non-negligible spins lies around between $0.25$--$0.71$ with a median of $0.46$, which is significantly higher compared to the fraction of our models. 
We note that this is a first order comparison, to properly compare our simulations with observations we need to take into account underlying assumptions such as our threshold spin of 0.05 and the priors used by \citet{Galaudage:2021rkt}.

\begin{figure*}
    \centering
    \includegraphics[width=1\textwidth]{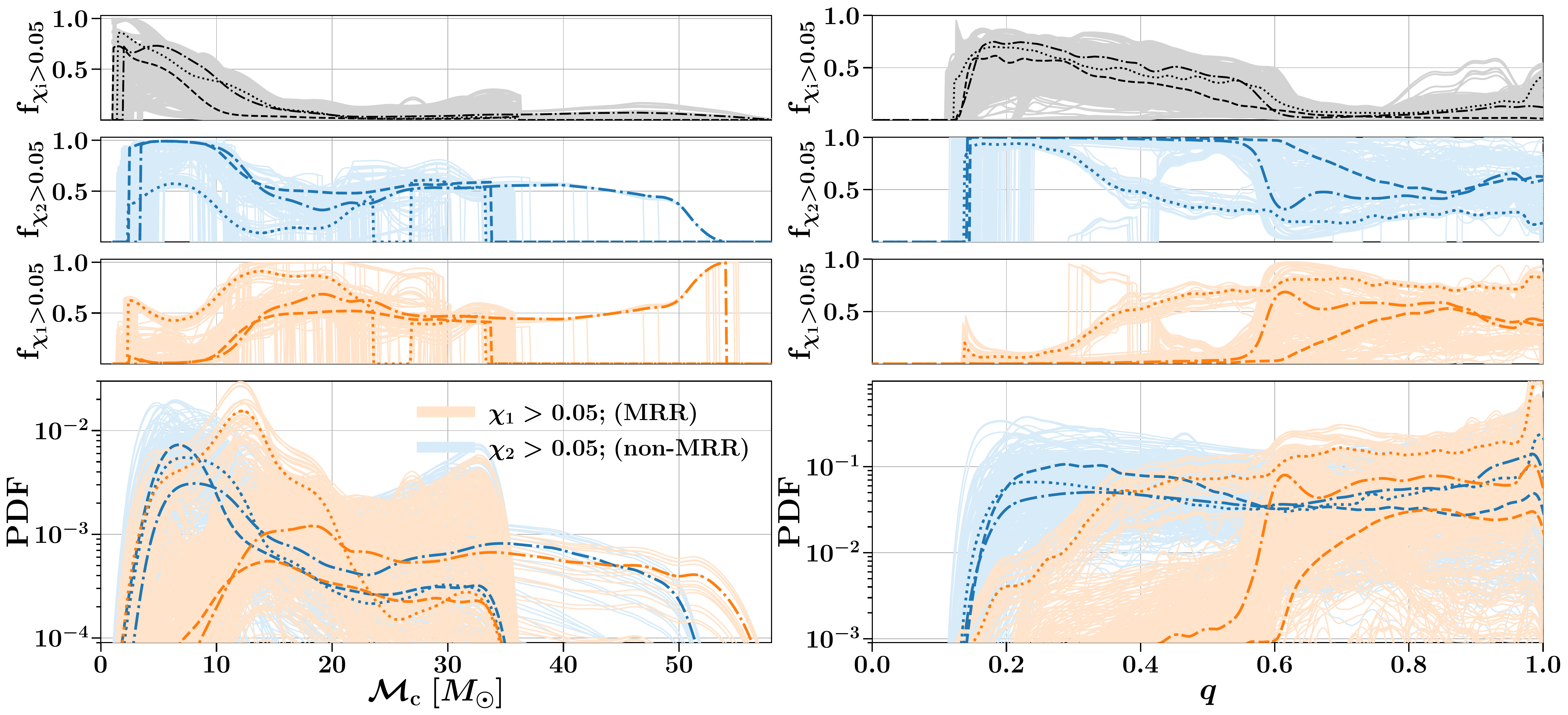}
    \caption{Chirp mass (left panels) and mass ratio (right panels) distributions split up in the subset of systems where the more massive black hole has non-negligible spin ($\chiBHone>0.05$; MRR) versus where the less massive black hole has non-negligible spin ($\chiBHtwo >0.05$; non-MRR). 
    Panels and labels are the same as in Fig.~\ref{fig:MRR-vs-not-MRR-BBH-properties}, but a top panel is added showing the overall fraction of binary black holes with one spinning component ($\chiBH > 0.05$). 
    \href{https://github.com/FloorBroekgaarden/MRR_Project/blob/main/Figure3_and_Figure5_BBH_properties_MRR_and_nonMRR_Distribution_plots/super_spins_split_2.pdf}{\faFileImage} \href{https://github.com/FloorBroekgaarden/MRR_Project/blob/main/Figure3_and_Figure5_BBH_properties_MRR_and_nonMRR_Distribution_plots/Make_Figure_3_and_5_Distribution_Plots.ipynb}{\faBook} 
    }
    \label{fig:properties_spinning}
\end{figure*}

Second, we find that the percentage among the systems with $\chiBH>0.05$ for which the more (less) massive black hole among the pair is spinning, i.e., $\chiBHone > 0.05$ ($\chiBHtwo > 0.05$), varies greatly in our models. 
For the more (less) massive black hole, these ranges span $0$--$60\%$ ($40$--$100\%$) and $0$--$80\%$ ($20$--$100\%$) for the astrophysical and detectable populations, respectively.
The largest fraction of binary black holes with $\chiBHone>0.05$ are presented in the models with the optimistic common envelope assumption (models F and K), where a significant fraction of the MRR binaries form through the classic common-envelope channel (Fig.~\ref{fig:FractionMRRbinaryBHs}).
This produces a relative large fraction of binary black holes with a $\chiBH>0.05$ as common-envelope evolution typically shrinks the binary to shorter orbital periods compared to stable mass transfer, allowing for more efficient tidal spin up.  
Another model assumption leading to a large fraction of systems having $\chiBHone > 0.05$ is model J which assumes an extremely high common-envelope efficiency ($\alpha=10$), which also results in tighter binaries.

Third, we notice that both the stellar evolution and \SFRD variations can significantly impact the expected fraction of binary black holes with $\chiBH >0.05$ as well as the fraction of systems with $\chiBHone>0.05$ or $\chiBHtwo>0.05$. 
For example, within the stellar evolution model with the optimistic common-envelope assumption (K) the fraction of systems with $\chiBH >0.05$  varies from $0.2$--$0.4$ and $0.05$--$0.25$ depending on the chosen \SFRD model for the astrophysical and detectable population, respectively.  
The sensitivity of our model realizations to the spin fraction, as well as which black hole is spinning, indicates that improved measurements of spin fractions inferred from future observations could aid in constraining parameters in the stellar and cosmic evolution assumptions within the isolated binary evolution channel.

\subsection{Properties of spinning binary black holes}
\label{subsec:propertie-of-spinning-binary-black-holes}
In the previous section we showed that both the MRR and non-MRR channel can significantly contribute to a sub-population of binary black holes with non-negligible spin and that they are identifiable from observations by having non-negligible $\chiBHone$ and $\chiBHtwo$, respectively. To further investigate the properties of these populations we show the chirp mass and mass ratio distributions for the subset of binary black holes with non-negligible spin ($\chiBH > 0.05$) for both the subsets for which the spinning black hole is the more massive black hole ($\chiBHone>0.05$) or the less massive black hole  ($\chiBHtwo>0.05$) in Fig.~\ref{fig:properties_spinning}.  Overall, we find that as a function of the black hole properties our models allow a wide range of contributions from both channels cf. Fig.~\ref{fig:frac_spinning}. We mention several visible features.

First, as can be seen from the top row panels in Fig.~\ref{fig:properties_spinning}, our models allow the majority of binary black holes to have non-negligible spin for systems with $\Mchirp \lesssim 10\Msun$ and $\q \lesssim 0.4$. For binary black holes with $\Mchirp \gtrsim 10\Msun$ or $\q \gtrsim 0.4$, on the other hand, the fraction of binary black hole mergers that contain a black hole with spin $\chiBH > 0.05$ rapidly declines. This is because these higher chirp mass or more equal mass ratio systems contain more massive second-born black holes. These more massive second-born black holes form from a black hole--Wolf Rayet binary with a more massive Wolf-Rayet star and a wider orbit, which disfavor tidal spin up in our model from  \citet{Bavera:2021evk}, see \S\ref{method:black_hole_spin_modelling}.

Second, the panels in the second and third row in Fig.~\ref{fig:properties_spinning} show the fraction of binary black holes with $\chiBH>0.05$ where we expect the less massive and more massive black hole to be spinning, respectively. 
The left panels indicate that we expect for systems with chirp mass $\Mchirp \lesssim 10\Msun$ in most models $\gtrsim 50\%$ ($\lesssim 50\%$) of the binary black holes with non-negligible spin to have the less (more) massive black hole $m_2$ ($m_1$) spinning, i.e., $\chiBHtwo > 0.05$ ($\chiBHone > 0.05$). On the other hand,  this contribution varies between $0\%$--$50\%$ ($50\%$--$100\%$) for systems with chirp mass $\gtrsim 10\Msun$. In all models, except our models that does not assume pair-instability supernovae to occur (O), we do not create binary black holes with $\Mchirp \gtrsim 35\Msun$, which creates the steep decline in Fig.~\ref{fig:properties_spinning} around this value. In models L we find that for the spinning binary black holes with $\Mchirp \gtrsim 35$ we expect approximately equally that the more or less massive black hole is spinning. 
For the mass ratio (right panels) we find that only the models with the optimistic common-envelope assumption (F and K) allow a large fraction of MRR binary black holes with non-negligible spin to significantly contribute at $q \lesssim 0.4$, though these systems remain rare in the overall population (Fig.~\ref{fig:MRR-vs-not-MRR-BBH-properties}). For $q \gtrsim 0.4$ we find that both MRR and non-MRR binaries can significantly contribute to binary black hole population with a spinning black hole, indicating that we expect to find both systems with $\chi_1>0.05$ and $\chi_2 > 0.05$, where the relative contributions depend on the model realization. 


Finally, the bottom panels in Fig.~\ref{fig:properties_spinning}, show the normalized distribution functions of the chirp mass and mass ratio for the subset of binary black holes with a non-negligible spinning black hole.  We find that the chirp mass distributions roughly follows the shape in Fig.~\ref{fig:MRR-vs-not-MRR-BBH-properties}, except that the contribution of non-MRR binaries are much higher in Fig.~\ref{fig:properties_spinning} as a result from the bias towards non-MRR systems for binary black holes with a non-negligible spin. For the mass ratio we find a much flatter distribution in  Fig.~\ref{fig:properties_spinning} (compared to Fig.~\ref{fig:MRR-vs-not-MRR-BBH-properties}) as a result that second-born black holes with lower masses are more likely to be spinning, leading to a boost of more extreme mass ratio systems for the non-MRR channel and an overall decrease of the contribution of $q\sim 1$ systems to the spinning population.

\begin{figure*}
    \centering
    \includegraphics[width=1\textwidth]{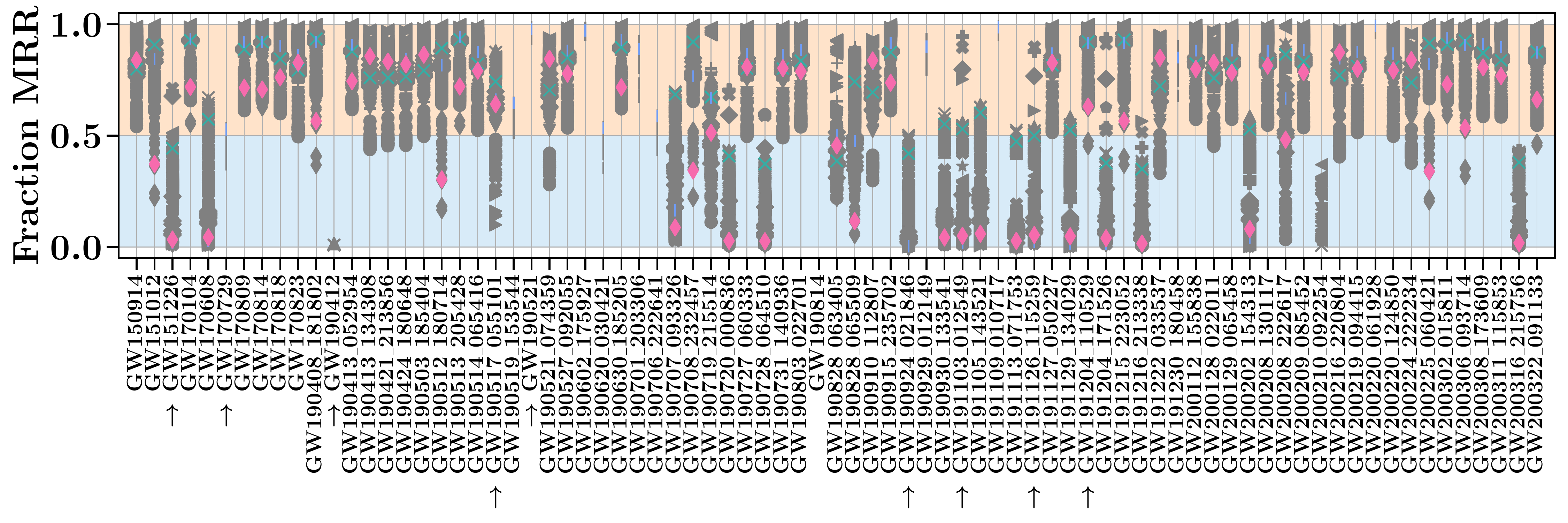}
    \caption{Fraction of binary black hole systems that match the inferred $90\%$ credible contours  that are \ac{MRR} (versus non-MRR) for the gravitational-wave events from GWTC-1, GWTC-2, and GWTC-3. Arrows indicate the observations that are discussed in the text.  Three realizations, `K123', `O312' and `T231' (see \href{https://github.com/FloorBroekgaarden/MRR_Project/blob/main/README.md}{Github}) are highlighted with a cross, vertical line, and diamond marker, respectively.
    \href{https://github.com/FloorBroekgaarden/MRR_Project/blob/main/Figure6_GW_Ratio_MRR_vs_non_MRR/Rate_allGWs_MRR.pdf}{\faFileImage} \href{https://github.com/FloorBroekgaarden/MRR_Project/blob/main/Figure6_GW_Ratio_MRR_vs_non_MRR/fraction_MRR_for_GW_CI.ipynb}{\faBook} 
    }
    \label{fig:MRR-fraction-per_GW-observation}
\end{figure*}
%

\section{Discussion}
\label{sec:discussion}

In this section we discuss some of the implications of our models for interpreting gravitational-wave observations.

\subsection{Comparison with observations}
\label{subsec:GW151226_and_friends}

The motivation for this work is to explain how MRR affects our interpretation of gravitational-wave signals and to make testable predictions about MRR for gravitational-wave astronomy. We noted in \S~\ref{subsec:propertie-of-spinning-binary-black-holes} that among the binary black hole population with non-negligible spin, for MRR systems $\chiBHone$ is spinning (and \chiBHtwo is zero) whereas for non-MRR  $\chiBHtwo$ is spinning (and \chiBHone is zero). We also found that among the non-negligible spin population non-MRR binaries are more common (Fig.~\ref{fig:frac_spinning}),  despite MRR binary black holes being more numerous in the overall population (Fig.~\ref{fig:FractionMRRbinaryBHs}). 
It is possible to measure the magnitudes and directions of black hole spins from their imprint on the observed gravitational waveform.
In most cases, the individual spin components are not well constrained, and often a mass weighted combination of the spins known as the effective inspiral spin parameter $\chi_\mathrm{eff}$ is the best constrained spin parameter. 

\begin{figure*}
    \centering
    \includegraphics[width=1\textwidth]{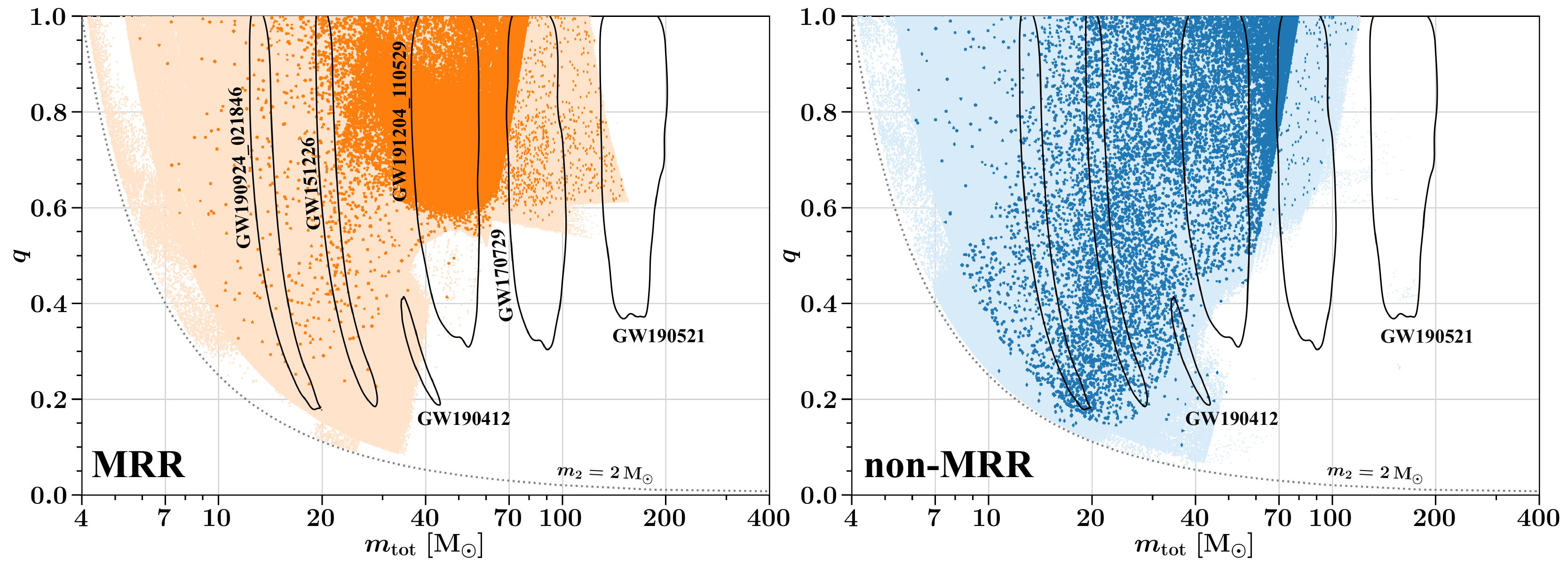}
    \caption{Total mass and mass ratio for MRR (left panel) and non-MRR (right panel) binary black holes in all our combined 560 model realizations. We draw from each model realization 100 binary black holes  between the combined MRR and non-MRR population and show these draws with a darker hue to visualize the density of points. 
    The $90\%$ credible contours for six gravitational-wave events that are discussed in the text are also shown.
    \href{https://github.com/FloorBroekgaarden/MRR_Project/blob/main/Figure_7_and_8_and_9_and_10_Discussion_2D_slices/Mtot_vs_q_all_models_all_pointsallNonMRR.png}{\faFileImage} \href{https://github.com/FloorBroekgaarden/MRR_Project/blob/main/Figure_7_and_8_and_9_and_10_Discussion_2D_slices/Make_2D_scatter_Distributions.ipynb}{\faBook} 
    }
    \label{fig:GW-observations-Mtot-versus-q}
\end{figure*}

\begin{figure*}
    \centering
    \includegraphics[width=1\textwidth]{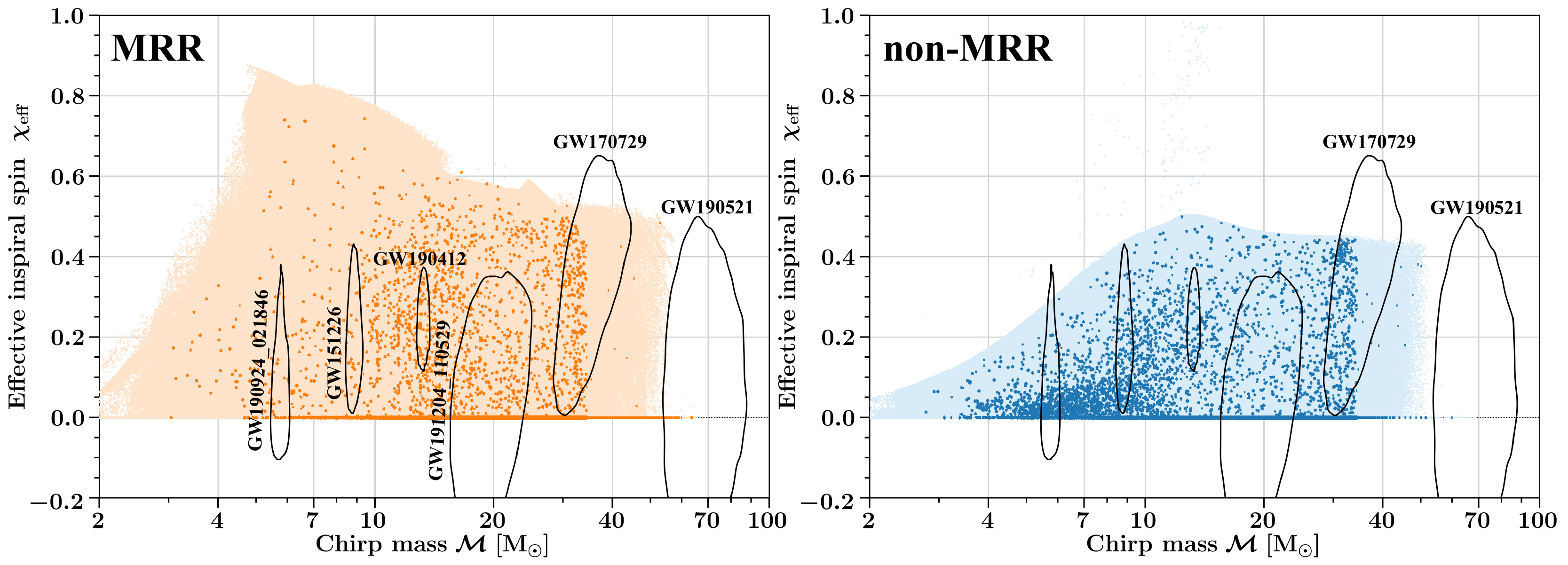}
    \caption{Same as Fig.~\ref{fig:GW-observations-Mtot-versus-q} for the chirp mass and effective spin.  
    \href{https://github.com/FloorBroekgaarden/MRR_Project/blob/main/Figure_7_and_8_and_9_and_10_Discussion_2D_slices/ChiEff_vs_q_all_models_all_pointsallNonMRR.png}{\faFileImage} \href{https://github.com/FloorBroekgaarden/MRR_Project/blob/main/Figure_7_and_8_and_9_and_10_Discussion_2D_slices/Make_2D_scatter_Distributions.ipynb}{\faBook} 
    }
    \label{fig:GW-observations-Chirp-mass_vs_chi_effective}
\end{figure*}

To give the reader an intuitive feeling for which of the  gravitational-wave detections are MRR or non-MRR candidates, we show in  Figure~\ref{fig:MRR-fraction-per_GW-observation} the MRR probability for each of the 79 binary black hole sources from the GWTC-1, GWTC-2, and GWTC-3 catalogs \citep{2019PhRvX...9c1040A,Abbott:2020niy,LIGOScientific:2021djp}. To do this we first select for each gravitational wave detection all binary black hole samples in our model realizations that have total mass, chirp mass, mass ratio, individual masses and effective spins that fall inside the inferred $90\%$ credible intervals from LVK. We then determine the ratio of the weight of the MRR versus  non-MRR samples that match each gravitational-wave credible interval.\footnote{We note that this method of calculating the ratio between MRR and non-MRR is adhoc and does not take into account the more complex weighting of the posterior distributions (or likelihoods) for the gravitational-wave events detected by LVK. We do this to get a rough feeling of whether MRR or non-MRR binaries dominate in a box in the parameter space around the gravitational-wave events. A more detailed study is outside of the scope of this work. We did test that our choice of $90\%$ credible intervals (i.e., which parameters we choose to mask in) did not significantly impact our results.} In addition, in Fig.~\ref{fig:GW-observations-Mtot-versus-q} and Fig.~\ref{fig:GW-observations-Chirp-mass_vs_chi_effective} we show two-dimensional slices of the MRR and non-MRR binary black hole properties overlaid with the credible contours of several gravitational-wave observations.   We discuss our results in light of some of the most relevant events below.

Overall, we find that the likelihood that a gravitational-wave source is MRR is strongly model dependent, as can be seen by the large variation per event in Fig.~\ref{fig:MRR-fraction-per_GW-observation}. This results from the overlap in binary black hole properties between MRR and non-MRR in our simulations (as shown in Fig.~\ref{fig:GW-observations-Mtot-versus-q} and Fig.~\ref{fig:GW-observations-Chirp-mass_vs_chi_effective}), the broad variety in MRR contributions to the detectable population already shown in Fig.~\ref{fig:FractionMRRbinaryBHs} and Fig.~\ref{fig:frac_spinning}, as well as the large uncertainties in many of the inferred credible intervals. GW191204$\_$110529 is an example of a system with one of the highest MRR fractions ($\gtrsim 0.5$) in Fig.~\ref{fig:MRR-fraction-per_GW-observation} for all model realizations, making it a likely MRR candidate due to its total mass around $m_{\rm{tot}}\sim 50\Msun$. 
On the other hand,  GW190924$\_$021846 is a system with one of the lowest MRR fractions ($\lesssim 0.5$) making it a likely non-MRR candidate due to its lower total mass $m_{\rm{tot}}\sim 14\Msun$ (see Fig.~\ref{fig:GW-observations-Mtot-versus-q}).

The earliest example of a binary black hole showing evidence for a highly spinning component was GW151226 \citep{LIGOScientific:2016sjg}, the second binary black hole merger to be observed.
\citet{LIGOScientific:2016sjg} found that GW151226 originated from a binary black hole merger with a total mass of around $22$\,M$_\odot$, and at least one of the component black holes had a spin greater than 0.2.
\citet{LIGOScientific:2018mvr} find GW151226 to have a mass of the more massive black hole of $m_{1} = 13.7^{+8.8}_{-3.2}$\,M$_\odot$, a mass of the less massive black hole of 
$m_{2} = 7.7^{+2.2}_{-2.5}$\,M$_\odot$,  and an effective spin of $\chi_\mathrm{eff} = 0.18^{+0.20}_{-0.12}$. 
Its individual spin components are less well constrained to $\chi_1 =0.69_{-0.55}^{+0.28}$ and $\chi_2 =0.50_{-0.45}^{+0.28}$ (cf. also \citealt{Biscoveanu:2020are}).
Subsequent reanalyses have utilised gravitational waveforms including the contributions of higher order modes \citep{Chia:2021mxq,Nitz:2021uxj}.
These analyses found evidence that GW151226 may be have more asymmetric masses than previously thought.
However, \citet{Mateu-Lucena:2021siq} argue that these results arise primarily from issues with the sampling of the posterior distribution, though \citet{Vajpeyi:2022dvs} deliver a mixed verdict: finding roughly equal support for the low-$q$ and high-$q$ hypotheses.
In addition to the technical challenge of sampling the posterior distribution, constraints on black hole spins can also vary significantly depending on the choice of prior assumptions \citep{Vitale:2017cfs,Zevin:2020gxf}.
In the full sample of observed binary black holes, there are now several examples of binaries with similar parameters to GW151226, such as the just marginal event GW191103\_012549 and GW191126\_115259  \citep{LIGOScientific:2021djp}.
The relatively high spins and low masses of GW151226-like events are broadly consistent with our models and can be explained by both a MRR and non-MRR formation, with the non-MRR being (strongly) favored with MRR fraction of $\lesssim 0.5$ for most of our model realizations in Fig.~\ref{fig:MRR-fraction-per_GW-observation} for  GW151226, GW191103\_012549, and GW191126\_115259 as can also be seen from the location of the GW151226 contour in   Fig.~\ref{fig:GW-observations-Mtot-versus-q} and Fig.~\ref{fig:GW-observations-Chirp-mass_vs_chi_effective}.

There also appears to be a class of more massive binary black holes with large positive effective spins. 
This includes the event GW170729 from GWTC-1 \citep[][]{LIGOScientific:2018mvr}. 
The mass properties of these events are only consistent with systems in our simulations in the models that do not assume a pair-instability mass gap (models O), as can be seen from Fig.~\ref{fig:MRR-vs-not-MRR-BBH-properties}.
A particularly intriguing event from GWTC-2 is GW190517\_055101  \citep[][]{Abbott:2020niy}, which has a total mass of $63.5^{+9.6}_{-9.6}$\,M$_\odot$, and an effective spin of $\chi_\mathrm{eff} = 0.52^{+0.19}_{-0.19}$. 
This matches events for both MRR and non-MRR in our simulations.

In binary black holes with a mass ratio that can be confidently measured away from unity, the dominant contribution to the effective spin comes from the spin of the more massive black hole, $\chi_{1}$, allowing it to be  precisely constrained. 
One such event is GW190412 \citep{LIGOScientific:2020stg}, inferred to be from a binary black hole merger with component masses of $m_1 = 30.1^{+4.6}_{-5.3}$\,M$_\odot$ and $m_2 = 8.3^{+1.6}_{-0.9}$\,M$_\odot$, 
constraining the mass ratio $q = m_2 / m_1 = 0.28^{+0.12}_{-0.07}$
(all values quoted at 90\% confidence), confidently excluding an equal mass binary.
For GW190412, the effective inspiral spin was measured to be $\chi_\mathrm{eff} = 0.25^{+0.08}_{-0.11}$, constraining the spin of the more massive black hole $\chi_1 = 0.44^{+0.16}_{-0.22}$, suggesting the presence of a rapidly rotating more massive black hole.

\citet{Mandel:2020lhv} argued for an alternative interpretation in which the more massive black hole in GW190412 was non-spinning, but the less massive black hole was rapidly spinning.
This was motivated by models of isolated binary evolution in which the second-born black hole can be tidally spun up in a tight binary, as described above.

\citet{Zevin:2020gxf} later challenged the interpretation of \citet{Mandel:2020lhv}.
They systematically analysed the gravitational-wave data for GW190412 under an array of different prior assumptions regarding the black hole spins. 
They examined which spin priors were preferred by the data by calculating the Bayesian evidence under each prior. 
They showed that a binary configuration allowing for a spinning more massive black hole was mildly preferred over the case in which the more massive black hole was non-spinning, in agreement with \citet{LIGOScientific:2020stg}, and disfavouring the prior preferred by \citet{Mandel:2020lhv}.

In terms of the chirp mass, for binary black holes with chirp masses similar to that of GW190412 ($\sim 13$\,M$_\odot$) mass ratio reversal is common ($f_\mathrm{MRR} \gtrsim 30\%$, Fig.~\ref{fig:MRR-vs-not-MRR-BBH-properties}).
However, in our models, merging binary black holes with mass ratios $q < 0.5$ are uncommon at any chirp mass (cf. Fig.~\ref{fig:MRR-vs-not-MRR-BBH-properties}), implying that an isolated binary evolution origin for GW190412 is unlikely (though see \citealp{Olejak:2020oel}, who find that $\sim 10\%$ of merging binary black holes have a mass ratio $q < 0.4$ in their model). Indeed, only in our simulations that assume the optimistic common-envelope prescription (models F and K) do we find systems matching GW190412. 
If GW190412 was formed through isolated binary evolution, our models suggest that mass ratio reversal is unlikely (\pMRR{}\,$\lesssim 10^{-2}$) for binary black holes with such asymmetric masses (Fig.~\ref{fig:MRR-vs-not-MRR-BBH-properties}).
This supports the arguments of \citet{Mandel:2020lhv}, and makes the findings of a highly spinning more massive black hole \citep{LIGOScientific:2020stg,Zevin:2020gxf} difficult to understand through this channel. 
Several alternate explanations for the origin of GW190412 have been proposed, such as dynamical formation in a dense star cluster \citep{DiCarlo:2020lfa} including the possibility that the more massive black hole is the product of a previous black hole merger \citep{Rodriguez:2020viw,Gerosa:2020bjb,Liu:2020gif}.

Other examples of similar binaries may include GW190403\_051519  \citep{LIGOScientific:2021usb,Qin:2021vle}\footnote{GW190403\_051519  is not shown in Fig.~\ref{fig:MRR-fraction-per_GW-observation} as it is an event from the GWTC-2.1 catalog with a false alarm rate below $1 \rm{yr}^{-1}$, which we did not include.}, a binary black hole with a mass ratio of $q = 0.25^{+0.54}_{-0.11}$
and a more massive black hole spin of $\chi_{1} = 0.92^{+0.07}_{-0.22}$. 
\citet{Qin:2021vle} argued that the observation of such rapidly spinning black holes may indicate that the efficiency of angular momentum transport in massive stars is much weaker than otherwise argued.
However, the more massive component of GW190403\_051519, with a mass $m_1 = 88.0^{+28.2}_{-32.9}$\,\Msun{}, lies in the pair-instability mass gap, which is not expected to be populated in the isolated binary evolution channel \citep{1964ApJS....9..201F,2019ApJ...887...53F,Stevenson:2019rcw}, and which only our model without a pair-instability supernova implemented (model O) produces.
This may suggest that a second generation black hole is a plausible explanation for the more massive hole in this system. 
The system GW190521 \citep[][]{LIGOScientific:2020iuh,LIGOScientific:2020ufj} is very similar and we find that only our model without  pair-instability supernovae implemented produces such systems, in which case it is likely a MRR system (Fig.~\ref{fig:MRR-fraction-per_GW-observation}).

\begin{figure*}
    \centering
    \includegraphics[width=1\textwidth]{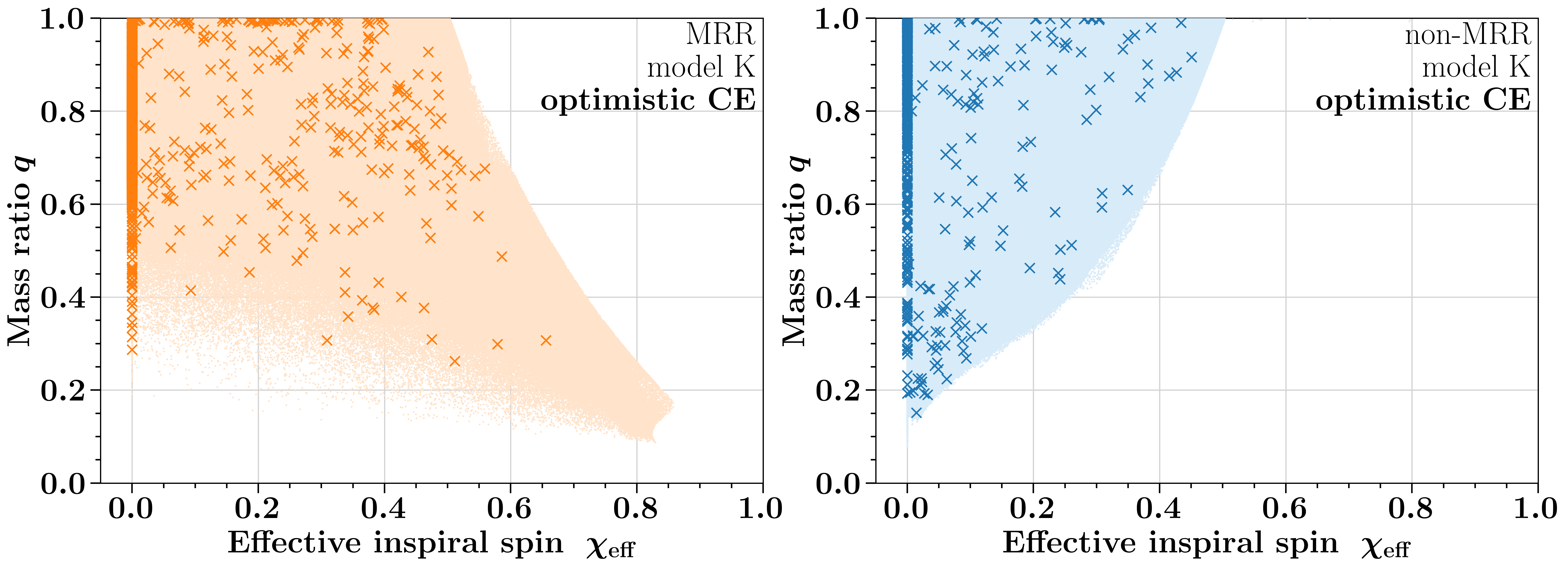}
     \caption{Same as 
    Fig.~\ref{fig:GW-observations-Mtot-versus-q} for the effective spin and mass ratio for our model with the optimistic common-envelope prescription (model K). 
    There is an anti-correlation present in the MRR binaries (left panel), but the opposite behavior is visible in the non-MRR binaries (right panel). For each of the 28 star formation history combinations we draw 1000 samples (combined over MRR and non-MRR) that we highlight with larger markers.  
    The other stellar evolution models are shown on \href{https://github.com/FloorBroekgaarden/MRR_Project/tree/main/Figure_7_and_8_and_9_and_10_Discussion_2D_slices/individual_ChiEff_vs_q}{GitHub}. 
    \href{https://github.com/FloorBroekgaarden/MRR_Project/blob/main/Figure_7_and_8_and_9_and_10_Discussion_2D_slices/individual_ChiEff_vs_q/ChiEff_vs_q_all_models_all_points_K.png}{\faFileImage} \href{https://github.com/FloorBroekgaarden/MRR_Project/blob/main/Figure_7_and_8_and_9_and_10_Discussion_2D_slices/Make_2D_scatter_Distributions.ipynb}{\faBook} 
    }
    \label{fig:chi_effective_q_correlation-plot}
\end{figure*}

In addition, recently it has been argued that the binary black hole observations by LVK suggest a possible anti-correlation between the  mass ratio $q$ and the effective inspiral spin parameter  $\chi_{\rm{eff}}$, such that more extreme mass ratios have higher effective spins \citep{LIGOScientific:2021djp,2021ApJ...922L...5C}. We find hints of a similar correlation for the MRR binary black holes in most of our model realizations, particularly those that assume the optimistic common-envelope prescription as shown in Fig.~\ref{fig:chi_effective_q_correlation-plot} (though our overall rate of extreme mass ratio events remains low). However, the non-MRR binary black holes in the simulations, that typically dominate the systems with non-negligible spin, do not show this correlation. This is because the effective spin is a mass weighted spin parameter such that for extreme mass ratios it is dominated by the spin of the most massive black hole, \chiBHone. In the MRR case $\chiBHone$ is non-negligible and \chiBHtwo is zero, which can lead to high effective spins for extreme mass ratios similar to the observed anti-correlation. On the other hand, in non-MRR systems the \chiBHtwo is non-negligble and \chiBHone is zero, resulting in lower spins for extreme mass ratios and the opposite correlation.

\begin{figure*}
    \centering
    \includegraphics[width=1\textwidth]{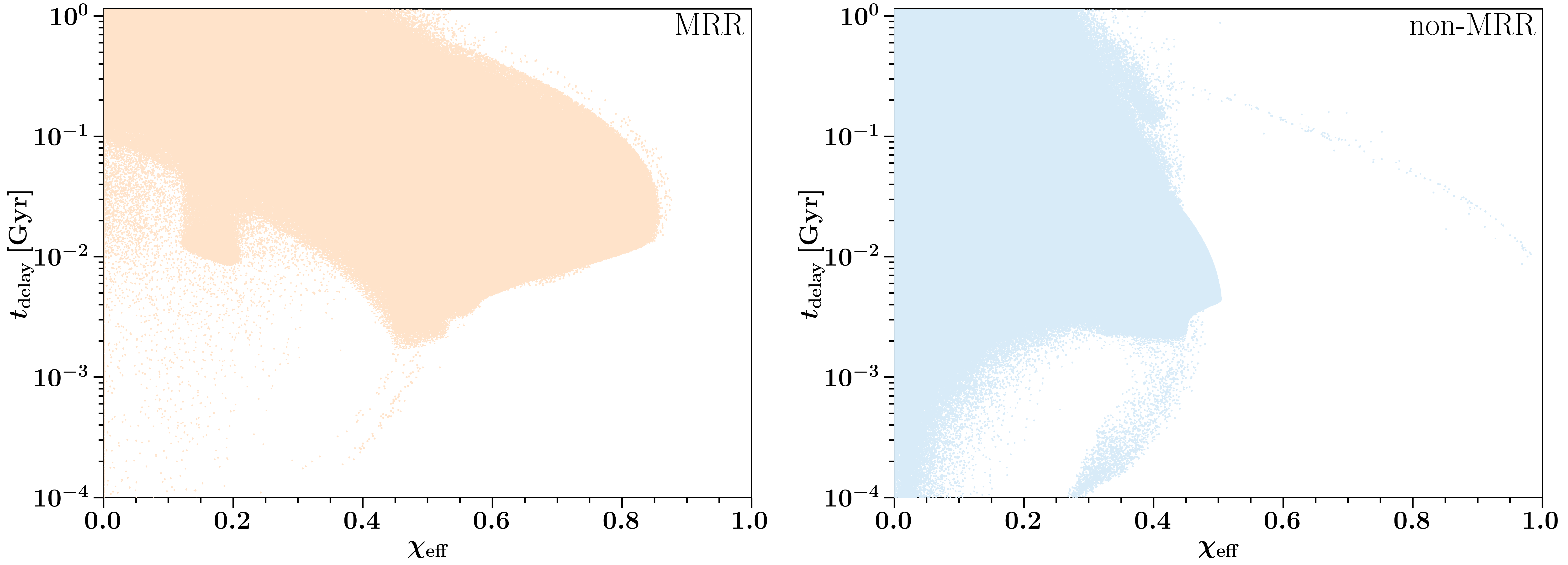}
     \caption{Effective spin and delay time for MRR (left) and non-MRR (right) binary black holes in our 20 stellar evolution simulations. Convolving the delay times with a metallicity-specific star formation history (\SFRD) provides the binary black holes  as a function of redshift. 
    There is an anti-correlation present where binary black holes with larger effective spins have shorter delay times, which is particularly visible in the MRR panel.
    We provide the same figure for each model separately at \href{https://github.com/FloorBroekgaarden/MRR_Project/tree/main/Figure_7_and_8_and_9_and_10_Discussion_2D_slices/chieff_tdelay_per_model}{GitHub}. 
    \href{https://github.com/FloorBroekgaarden/MRR_Project/blob/main/Figure_7_and_8_and_9_and_10_Discussion_2D_slices/ChiEff_vs_tdelay_2Dplot_priorWeighted_not_drawing.png}{\faFileImage} \href{https://github.com/FloorBroekgaarden/MRR_Project/blob/main/Figure_7_and_8_and_9_and_10_Discussion_2D_slices/Make_2D_scatter_Distributions.ipynb}{\faBook} 
    }
    \label{fig:chi_effective_delay-time_correlation-plot}
\end{figure*}

Last, recently several authors have suggested the existence of redshift-evolution in the binary black hole (effective) spin distribution: finding at higher redshifts both a larger fraction of binary black holes with non-negligible spin and a broadening in the black hole spin magnitude distribution \citet{Bavera:2022spinredshift,Biscoveanu:2022}. This positive spin-redshift correlation matches the expected behavior in our model assumptions. In our simulations, black holes dominantly spin up through tides  \citep[cf.][]{Bavera:2021evk} such that binaries with smaller separations at the black hole-Wolf Rayet phase are more commonly spun up and have stronger black hole spins. At the same time, the shorter separations also result in significantly smaller delay times between the formation of the binary black hole and its merger \citep{1964PhRv..136.1224P}. This leads to binary black hole systems with shorter delay times to also have higher spins, as shown for our simulations in Figure~\ref{fig:chi_effective_delay-time_correlation-plot}. 
At higher redshifts the binary black hole population has a larger contribution from systems with shorter delay times (which have already merged at lower redshifts and cannot contribute to this population). Together, the above leads to a redshift evolution in the spin distribution of binary black hole systems.  To understand the redshift behavior in more detail it is important to also include many other uncertainties in the assumptions of the spins, stellar evolution and star-formation history, which is outside of the scope of this study.

\subsection{Comparison with other formation channels}
\label{subsec:comparison_other_channels}

Our work studied the isolated binary evolution channel, but other channels to form binary black hole mergers have been proposed in the literature (see \citealp{Mandel:2022hfr} for a recent review). 
We compare our results to the main alternative formation channels below.

First, we did not consider the chemically homogeneous evolution pathway \citep{Mandel:2015qlu,Marchant:2016wow,Riley:2020btf}. 
Binaries formed through chemically homogeneous evolution are expected to have high black hole masses around $M \gtrsim 20\Msun$, mass ratios close to unity, and both components may be rapidly spinning \citep{Marchant:2016wow}. 
Among the population of binary black holes with non-negligible spin, a system might thus be particularly distinguishable from the systems studied in this paper by having both non-negligible $\chiBHone$ and $\chiBHtwo$ in combination with $q\sim 1$.  Typically, the contribution from chemically homogeneous evolution to the binary black hole population is expected to be small \citep[e.g.][]{MandelBroekgaarden:2021}.

Second, binary black holes may originate from dynamical interactions in dense stellar environments such as globular and open clusters.
Binary black holes formed in star clusters are expected to have an isotropic distribution of spins \citep[e.g.,][]{Rodriguez:2016vmx}.
Black holes born in star clusters likely have small spins ($\chi \sim 0$) for similar reasons to those discussed above, with the exception of a small fraction of black holes formed through previous black hole mergers, which are expected to have high spins, $\chi \sim 0.7$ \citep[e.g.,][]{Gerosa:2021mno}. This channel is expected to produce a significant fraction of binary black holes with misaligned spin, which would make such systems distinguishable from the systems expected in our study.

Alternatively, binary black holes assembled
in active galactic nuclei have been suggested to have spins with the more massive component spinning as a result from spherical and planar symmetry-breaking effects \citep[e.g.][]{McKernan:2021correlation}. They particularly showed that this can result in the mass ratio - $\chi_{\rm{effective}}$-correlation that LVK seems to observe,  though the contribution that this channel can have to the overall binary black hole population is still under debate.

\subsection{Neutron star-black hole binaries}
\label{subsec:NSBH}

In this paper we have focused on predictions for the masses and spins of binary black holes formed through isolated binary evolution involving a phase of mass ratio reversal.
Similar evolutionary scenarios to those shown in Fig.~\ref{fig:DetailedEvolutionMRR} can also occur at lower masses and produce neutron star-black hole binaries, where the (more massive) black hole is formed second \citep[][]{Chattopadhyay:2020lff,Broekgaarden:2021iew,Hu:2022ubh}. 
In the context of radio astronomy, this can be important for determining if the neutron star can be born first and recycled \citep{Chattopadhyay:2020lff}.
\citet{Broekgaarden:2021iew} calculated \pMRR{} for merging black hole neutron star binaries using the same set of models used here. 
They found that in most of these models, the neutron star is born first in less than 1\% of cases. 
In model D (high mass transfer efficiency) and H (optimistic CE) this rises to around 10\% of cases. 
The dramatic difference between neutron star black holes and binary black holes arises due to their different evolutionary pathways \citep{Broekgaarden:2021iew}.

This suggests that the black hole is almost always formed first in neutron star-black hole binaries, and thus, by the arguments given earlier, would be expected to be almost non-rotating. 
This supports the alternative interpretation of the neutron star black hole binary GW200115 \citep{LIGOScientific:2021qlt} by \citet{Mandel:2021ewy}, in which they argued that the black hole was consistent with being non-spinning.

\subsection{Eddington limited accretion}
\label{subsec:eddington_limit}

We have assumed the spin of the firstborn black hole to be zero.
This is justified by our assumption of Eddington-limited accretion onto compact objects, which limits the amount of mass accreted by black holes and therefore the accretion-induced spin-up.
It is possible that some mechanism exists to overcome the Eddington limit in these systems.
Allowing for super-Eddington accretion rates onto black holes could allow for greater amounts of accretion, and thus spin up of firstborn black holes.
Super-Eddington accretion is assumed by default in the BPASS models \citep[][]{BPASS:2017PASA}.
\citet{vanSon:2020zbk} investigated super-Eddington accretion using COMPAS.
They found that super-Eddington accretion did not strongly impact the occurrence of mass ratio reversal, but can significantly increase the mass of firstborn black holes (see their Fig.~1), although in many of these systems, the assumption of conservative mass transfer leads the orbit of the binary to widen significantly, such that these binaries no longer merge within the age of the Universe.
\citet{Bavera:2020uch} similarly found that allowing for highly super-Eddington accretion almost completely removes the contribution of the stable mass transfer channel, as conservative mass transfer is not as effective at reducing the orbital separation of the binary as the typically highly non-conservative mass transfer that occurs during Eddington limited accretion.
\citet{Zevin:2022wrw} also consider highly super-Eddington accretion and find that it can produce firstborn black holes with significantly higher spins \citep[see also][]{Shao:2022jzo}.

\subsection{Inferring the fraction of binary black holes having undergone mass ratio reversal from observations}
\label{subsec:phenom_model}

The predictions presented here can be used to inform the construction of phenomenological population models that take into account mass ratio reversal.
These phenomenological models can then be used to analyse the observed population of binary black hole mergers \citep[cf.][]{LIGOScientific:2021psn}.

Most phenomenological models for the spin distribution of binary black holes currently in use assume that the spins of both components are drawn from the same distribution \citep[][]{Talbot:2017yur,Wysocki:2018mpo,Galaudage:2021rkt,LIGOScientific:2021psn}, a decision motivated by the desire to limit the number of free parameters in the model, though see \citet{Biscoveanu:2020are} for an example of an analysis that does not make this assumption. 
A key feature of our model is that in some fraction of systems, the more massive black hole is spinning ($\chi_{1} > 0$), while the less massive black hole is non-spinning ($\chi_{2} = 0$).
Conversely, there is also a subpopulation where the less massive black hole is spinning ($\chi_{2} > 0$) and the more massive black hole is non-spinning ($\chi_{1} = 0$).
In the majority of systems ($\sim 80\%$) both black holes are non-spinning ($\chi_{1} = \chi_{2} = 0$).
Except for special cases, such as binary black holes formed through double core common envelope phases \citep[][]{Neijssel:2019,Broekgaarden:2021iew,Olejak:2021iux} or chemically homogeneous evolution \citep[][]{Marchant:2016wow}, we do not expect both black holes to have significant spins.
For the subpopulation of spinning black holes, the distribution is not robustly predicted by our models (see variation in the right hand panel of Fig.~\ref{fig:MRR-vs-not-MRR-BBH-properties}).
Regardless, one could use this figure to derive a flexible parametric model to describe the distribution of black hole spin magnitudes.
Such a phenomenological model should separately model the spin distribution of the more massive ($\chi_{1}$) and the less massive ($\chi_{2}$) black holes.

By determining the fraction of observed binary black holes in which the more massive is spinning ($\chi_{1} > 0$), one can place a lower limit on the fraction of binaries that have undergone mass ratio reversal ($f_\mathrm{MRR}$).
Similarly, a constraint on the fraction of systems in which the less massive is spinning ($\chi_{2} > 0$) allows one to place an upper limit on $f_\mathrm{MRR}$, though we expect that in practice this constraint will be weaker, as it is typically difficult to constrain the spin magnitude of the less massive black hole \citep[][]{LIGOScientific:2021djp}.
Unfortunately, it may be difficult to directly link the fraction of observed binary black holes with $\chi_{1} > 0$ to the true fraction of binaries that have undergone mass ratio reversal ($f_\mathrm{MRR}$), since we have shown that in many binary black holes formed through mass ratio reversal, both components are effectively non-spinning.
In our models, the former fraction is typically only around 5\%, while $f_\mathrm{MRR}$ is around 70\% (cf. Fig.~\ref{fig:frac_spinning}).
Binary population synthesis models such as those presented in this work may help bridge this gap.

Not only would constraining $f_\mathrm{MRR}$ observationally provide insight into the order in which the black holes formed, it is also closely linked to the fraction of binary black holes formed through only stable mass transfer, instead of common envelope evolution (cf. Fig.~\ref{fig:FractionMRRbinaryBHs}). 
Hence, a robust measurement of $f_\mathrm{MRR}$ could also provide insights into the formation pathways of binary black holes formed through massive binary evolution.

\subsection{Comparison with predictions from other population synthesis codes}
\label{subsec:compare_with_others}

We now briefly compare and contrast our key results with previous studies of binary black hole formation and mass ratio reversal.
One of the first population synthesis papers to study mass ratio reversal in binary black hole mergers was done by \citet{Gerosa:2013laa} who focused on the implications of mass ratio reversal in binary evolution for the precession of black hole spins in binary black holes using a suite of population synthesis models from \citet{Dominik:2012kk}.
These models predict $f_\mathrm{MRR}$ in the range 0.1--0.4, with a strong dependence on metallicity. 
Compared to \citet{Gerosa:2013laa}, we added new models and implement the contributions from different metallicities according to the metallicity-specific star formation history, accounting also for gravitational-wave selection effects. 

\citet{Gerosa:2018wbw} use an updated version of the models from \citet{Gerosa:2013laa}, accounting for gravitational-wave selection effects. 
For moderate black hole kicks, roughly one third of the binary black holes formed through common envelope evolution in their models undergo mass ratio reversal (see their Fig.~3 and associated discussion). 
At least part of the discrepancy between the results of \citet{Gerosa:2018wbw} and the present work can be attributed to different assumptions regarding the stability and efficiency of mass transfer (see \citealp{Belczynski:2021zaz} for further discussion).

\citet{Olejak:2021iux} studied the formation of binary black holes with high effective spin parameters through isolated binary evolution. 
They show an example of a mass ratio reversed binary black hole forming through stable mass transfer (see their Fig.~1) in which the more massive black hole is rapidly rotating as a result of tidal spin up, and also discuss that a fraction of detectable binary black holes is expected to have (significant) spin. 
However, \citet{Olejak:2021iux} do not quantify how common mass ratio reversal is in their models, nor how robust that prediction is under variations of the underlying binary evolution physics; these are the main goals of the present work.

Recently, \citet{Zevin:2022wrw} also studied mass ratio reversal in isolated binary evolution; their work complements our own and reinforces our findings.
In general, there is fairly good agreement between the two studies, with the largest differences arising due to different choices for the fiducial binary evolution model (e.g., the mass transfer efficiency). 
Our models explore a wider range of uncertainties in the cosmic star formation history than explored in \citet{Zevin:2022wrw}, whilst they explore a wider range of accretion efficiencies for compact objects, and explore models where multiple assumptions are varied from the fiducial assumption at once (something we have not done, with the exception of model F).
Across their suite of models, \citet{Zevin:2022wrw} find that more than 40\% of binary black holes have undergone mass ratio reversal for mass transfer accretion efficiencies $\beta > 0.5$ (with all other parameters at their fiducial values), with this fraction increasing with the assumed efficiency of mass transfer (as we also find in our models B, C and D, see Fig.~\ref{fig:FractionMRRbinaryBHs}).
Our default model translates to a relatively high mass transfer efficiency (as can be seen from Fig.~\ref{fig:DetailedEvolutionMRR}), and this likely explains why our results are towards the high end of those found by \citet{Zevin:2022wrw}.
They also find that in mass ratio reversed systems, the mass of the more massive black hole is rarely greater than twice the mass of the less massive black hole, again in broad agreement with our findings.

\section{Conclusions}
\label{sec:conclusions}

In this paper we investigated the occurrence of mass ratio reversal (MRR), where the more massive black hole is second-born, in the population of merging binary black holes originating from isolated massive binary stars. Using a large set of state-of-the-art population synthesis models from \citet{Broekgaarden:2021iew} we investigate the rate and properties of \ac{MRR} binary black hole mergers.
We discuss that MRR in binary black hole systems can have important observational consequences as the second-born black hole can be tidally spun up if the orbital period of the binary is short enough, whilst the first-born black hole is assumed to be non-spinning, which we model following  \citep[][]{Qin:2018vaa,Bavera:2019}.
This introduces the possibility of the more massive black hole in a merging binary black hole system being rapidly rotating. 
Our main findings are summarized below.
\begin{itemize}
    \item  We find that typically more than 70\% of observed binary black holes have undergone mass ratio reversal in most of our models (Fig.~\ref{fig:FractionMRRbinaryBHs}). I.e., we expect that for the majority of observed binary black hole mergers the most massive black hole component was born second. 
    \item  We find in  Fig.~\ref{fig:MRR-vs-not-MRR-BBH-properties} that MRR systems particularly dominate the binary black hole detections for larger chirp masses ($\gtrsim 10\Msun$) and larger mass ratios ($q\gtrsim 0.6$, with a peak of around $q \sim 0.7$). Non-MRR systems, on the other hand, dominate the binary black hole population for $\Mchirp < 10\Msun$ and $q<0.6$. 
    \item We show in Fig.~\ref{fig:frac_spinning} that our models predict up to  $25\%$ of detectable binary black holes  contain a component with non-negligible spin. 
    Among this non-negligible spin population, we find that the fraction of systems with non-negligible spin in which the more (less) massive black hole among the pair is spinning varies greatly in our models. 
    We expect the more (less) massive black hole to be spinning in $0\%$--$80\%$ ($20$--$100\%$) of the detectable non-negligible spin population.
    \item For the subset of spinning binary black holes, we investigated which component we expect to be spinning as a function of chirp mass and mass ratio. We found that overall our model realizations allow both the more massive (MRR) as well as the less massive black hole to be spinning (non-MRR) as a function of chirp mass and mass ratio values (Fig.~\ref{fig:properties_spinning}). For mass ratios below $q \lesssim 0.4$, however, we expect the less massive black hole to dominate the spin population in all models. 
    \item We discussed our results in context of the current population of binary black hole mergers in Section~\ref{subsec:GW151226_and_friends} and found that  candidates for mass-ratio reversed systems include GW191204$\_$110529  and similar binaries, whereas we find that binaries such as GW151226 and GW190412 \citep[][]{LIGOScientific:2020stg} are unlikely to have undergone mass ratio reversal based on their total mass and mass ratio properties (Fig.~\ref{fig:MRR-fraction-per_GW-observation},  Fig.~\ref{fig:GW-observations-Mtot-versus-q}, and Fig.~\ref{fig:GW-observations-Chirp-mass_vs_chi_effective}).
    \item We discussed in Section~\ref{subsec:GW151226_and_friends} that our MRR binaries do show hints of the observed anti-correlation between mass ratio and effective spin (Fig.~\ref{fig:chi_effective_q_correlation-plot}), but that in the overall population the binary black holes with non-negligible spins are dominated by non-MRR binaries. 
\end{itemize}

Overall, our results highlight that a significant fraction of the observed binary black hole mergers might have undergone mass ratio reversal such that the more massive black hole in the system formed second. We expect to identify these systems by mergers where the more massive black hole has non-negligible spin. Future observations will test our findings and help investigate stellar evolution assumptions.

\section*{Acknowledgements}
The authors are supported by the Australian Research Council (ARC) Centre of Excellence for Gravitational Wave Discovery OzGrav, through project number CE170100004.
SS is supported by the ARC Discovery Early Career Research Award (DE220100241). 
FSB acknowledges support by the Prins Bernard Cultuurfonds. 
The authors are thankful for useful discussions with Simone Bavera, Edo Berger, Sylvia Biscoveanu, Daria Gangardt, Davide Gerosa, Ilya Mandel, Matthew Mould, Lieke van Son, Salvo Vitale, and Mike Zevin.

\section*{Data Availability and Software} 
The authors made use of the binary black hole simulations from \citep{Broekgaarden:2021efa}, which are publicly available at  \citet[][]{Broekgaarden:2021-zenodo-BHBH}.  All code to reproduce the results and (additional) figures in this paper are publicly available on Github at \url{https://github.com/FloorBroekgaarden/MRR_Project} \href{https://github.com/FloorBroekgaarden/MRR_Project}{\faGithub}. In addition, a link to the specific jupyter notebook to recreate a figure is provided in each figure caption.

The authors made use of \textsc{Python} from the  Python Software Foundation. Python Language Reference, version 3.6. Available at \url{http://www.python.org} \citep{CS-R9526}. In addition the following Python packages were used: \textsc{matplotlib} \citep{2007CSE.....9...90H},  \textsc{NumPy} \citep{2020NumPy-Array}, \textsc{SciPy} \citep{2020SciPy-NMeth}, \texttt{ipython$/$jupyter} \citep{2007CSE.....9c..21P, kluyver2016jupyter}, \textsc{pandas} \citep{mckinney-proc-scipy-2010}, \textsc{Seaborn} \citep{waskom2020seaborn}, \textsc{Astropy} \citep{2018AJ....156..123A}  and   \href{https://docs.h5py.org/en/stable/}{\textsc{hdf5}} \citep{collette_python_hdf5_2014}. 

 In addition, we used the posterior samples for the events from the GWTC-1, GWTC-2, and GWTC-3 catalog provided by the Gravitational Wave Open Science Center (\url{https://www.gw-openscience.org/}), a service of LIGO Laboratory, the LIGO Scientific Collaboration and the Virgo Collaboration. LIGO Laboratory and Advanced LIGO are funded by the United States National Science Foundation (NSF) as well as the Science and Technology Facilities Council (STFC) of the United Kingdom, the Max-Planck-Society (MPS), and the State of Niedersachsen/Germany for support of the construction of Advanced LIGO and construction and operation of the GEO600 detector. Additional support for Advanced LIGO was provided by the Australian Research Council. Virgo is funded, through the European Gravitational Observatory (EGO), by the French Centre National de Recherche Scientifique (CNRS), the Italian Istituto Nazionale di Fisica Nucleare (INFN) and the Dutch Nikhef, with contributions by institutions from Belgium, Germany, Greece, Hungary, Ireland, Japan, Monaco, Poland, Portugal, Spain.
 This research has made use of NASA`s Astrophysics Data System Bibliographic Services.

\bibliography{refs} 

\providecommand{\noopsort}[1]{}
\begin{thebibliography}{}
\expandafter\ifx\csname natexlab\endcsname\relax\def\natexlab#1{#1}\fi
\providecommand{\url}[1]{\href{#1}{#1}}
\providecommand{\dodoi}[1]{doi:~\href{http://doi.org/#1}{\nolinkurl{#1}}}
\providecommand{\doeprint}[1]{\href{http://ascl.net/#1}{\nolinkurl{http://ascl.net/#1}}}
\providecommand{\doarXiv}[1]{\href{https://arxiv.org/abs/#1}{\nolinkurl{https://arxiv.org/abs/#1}}}

\bibitem[{Aasi {et~al.}(2015)}]{TheLIGOScientificDetector:2014jea}
Aasi, J., {et~al.} 2015, Class. Quant. Grav., 32, 074001,
  \dodoi{10.1088/0264-9381/32/7/074001}

\bibitem[{Abbott {et~al.}(2016)}]{LIGOScientific:2016sjg}
Abbott, B.~P., {et~al.} 2016, Phys. Rev. Lett., 116, 241103,
  \dodoi{10.1103/PhysRevLett.116.241103}

\bibitem[{Abbott {et~al.}(2019{\natexlab{a}})}]{LIGOScientific:2018mvr}
---. 2019{\natexlab{a}}, Phys. Rev. X, 9, 031040,
  \dodoi{10.1103/PhysRevX.9.031040}

\bibitem[{Abbott {et~al.}(2019{\natexlab{b}})}]{2019PhRvX...9c1040A}
Abbott, R., {et~al.} 2019{\natexlab{b}}, Physical Review X, 9, 031040,
  \dodoi{10.1103/PhysRevX.9.031040}

\bibitem[{Abbott {et~al.}(2020{\natexlab{a}})}]{LIGOScientific:2020stg}
---. 2020{\natexlab{a}}, Phys. Rev. D, 102, 043015,
  \dodoi{10.1103/PhysRevD.102.043015}

\bibitem[{Abbott {et~al.}(2020{\natexlab{b}})}]{LIGOScientific:2020iuh}
---. 2020{\natexlab{b}}, Phys. Rev. Lett., 125, 101102,
  \dodoi{10.1103/PhysRevLett.125.101102}

\bibitem[{Abbott {et~al.}(2020{\natexlab{c}})}]{LIGOScientific:2020ufj}
---. 2020{\natexlab{c}}, Astrophys. J. Lett., 900, L13,
  \dodoi{10.3847/2041-8213/aba493}

\bibitem[{Abbott {et~al.}(2021{\natexlab{a}})}]{Abbott:2020niy}
---. 2021{\natexlab{a}}, Physical Review X, 11, 021053,
  \dodoi{10.1103/PhysRevX.11.021053}

\bibitem[{Abbott {et~al.}(2021{\natexlab{b}})}]{LIGOScientific:2021djp}
---. 2021{\natexlab{b}}.
\newblock \doarXiv{2111.03606}

\bibitem[{Abbott {et~al.}(2021{\natexlab{c}})}]{LIGOScientific:2021usb}
---. 2021{\natexlab{c}}.
\newblock \doarXiv{2108.01045}

\bibitem[{Abbott {et~al.}(2021{\natexlab{d}})}]{LIGOScientific:2021qlt}
---. 2021{\natexlab{d}}, Astrophys. J. Lett., 915, L5,
  \dodoi{10.3847/2041-8213/ac082e}

\bibitem[{Abbott {et~al.}(2021{\natexlab{e}})}]{LIGOScientific:2021psn}
---. 2021{\natexlab{e}}.
\newblock \doarXiv{2111.03634}

\bibitem[{Acernese {et~al.}(2015)}]{TheVirgoDetector:2014hva}
Acernese, F., {et~al.} 2015, Class. Quant. Grav., 32, 024001,
  \dodoi{10.1088/0264-9381/32/2/024001}

\bibitem[{{Agrawal} {et~al.}(2020){Agrawal}, {Hurley}, {Stevenson},
  {Sz{\'e}csi}, \& {Flynn}}]{Agrawal:2020MNRAS}
{Agrawal}, P., {Hurley}, J., {Stevenson}, S., {Sz{\'e}csi}, D., \& {Flynn}, C.
  2020, \mnras, 497, 4549, \dodoi{10.1093/mnras/staa2264}

\bibitem[{Akutsu {et~al.}(2019)}]{KAGRA:2018plz}
Akutsu, T., {et~al.} 2019, Nature Astron., 3, 35,
  \dodoi{10.1038/s41550-018-0658-y}

\bibitem[{{Astropy Collaboration} {et~al.}(2018){Astropy Collaboration},
  {Price-Whelan}, {Sip{\H{o}}cz}, {G{\"u}nther}, {Lim}, {Crawford}, {Conseil},
  {Shupe}, {Craig}, {Dencheva}, {Ginsburg}, {Vand erPlas}, {Bradley},
  {P{\'e}rez-Su{\'a}rez}, {de Val-Borro}, {Aldcroft}, {Cruz}, {Robitaille},
  {Tollerud}, {Ardelean}, {Babej}, {Bach}, {Bachetti}, {Bakanov}, {Bamford},
  {Barentsen}, {Barmby}, {Baumbach}, {Berry}, {Biscani}, {Boquien}, {Bostroem},
  {Bouma}, {Brammer}, {Bray}, {Breytenbach}, {Buddelmeijer}, {Burke},
  {Calderone}, {Cano Rodr{\'\i}guez}, {Cara}, {Cardoso}, {Cheedella}, {Copin},
  {Corrales}, {Crichton}, {D'Avella}, {Deil}, {Depagne}, {Dietrich}, {Donath},
  {Droettboom}, {Earl}, {Erben}, {Fabbro}, {Ferreira}, {Finethy}, {Fox},
  {Garrison}, {Gibbons}, {Goldstein}, {Gommers}, {Greco}, {Greenfield},
  {Groener}, {Grollier}, {Hagen}, {Hirst}, {Homeier}, {Horton}, {Hosseinzadeh},
  {Hu}, {Hunkeler}, {Ivezi{\'c}}, {Jain}, {Jenness}, {Kanarek}, {Kendrew},
  {Kern}, {Kerzendorf}, {Khvalko}, {King}, {Kirkby}, {Kulkarni}, {Kumar},
  {Lee}, {Lenz}, {Littlefair}, {Ma}, {Macleod}, {Mastropietro}, {McCully},
  {Montagnac}, {Morris}, {Mueller}, {Mumford}, {Muna}, {Murphy}, {Nelson},
  {Nguyen}, {Ninan}, {N{\"o}the}, {Ogaz}, {Oh}, {Parejko}, {Parley}, {Pascual},
  {Patil}, {Patil}, {Plunkett}, {Prochaska}, {Rastogi}, {Reddy Janga},
  {Sabater}, {Sakurikar}, {Seifert}, {Sherbert}, {Sherwood-Taylor}, {Shih},
  {Sick}, {Silbiger}, {Singanamalla}, {Singer}, {Sladen}, {Sooley},
  {Sornarajah}, {Streicher}, {Teuben}, {Thomas}, {Tremblay}, {Turner},
  {Terr{\'o}n}, {van Kerkwijk}, {de la Vega}, {Watkins}, {Weaver}, {Whitmore},
  {Woillez}, {Zabalza}, \& {Astropy Contributors}}]{2018AJ....156..123A}
{Astropy Collaboration}, {Price-Whelan}, A.~M., {Sip{\H{o}}cz}, B.~M., {et~al.}
  2018, \aj, 156, 123, \dodoi{10.3847/1538-3881/aabc4f}

\bibitem[{{Bavera} {et~al.}(2022){Bavera}, {Fishbach}, {Zevin}, {Zapartas}, \&
  {Fragos}}]{Bavera:2022spinredshift}
{Bavera}, S.~S., {Fishbach}, M., {Zevin}, M., {Zapartas}, E., \& {Fragos}, T.
  2022, arXiv e-prints, arXiv:2204.02619.
\newblock \doarXiv{2204.02619}

\bibitem[{{Bavera} {et~al.}(2021){Bavera}, {Zevin}, \&
  {Fragos}}]{Bavera:2021evk}
{Bavera}, S.~S., {Zevin}, M., \& {Fragos}, T. 2021, Research Notes of the
  American Astronomical Society, 5, 127, \dodoi{10.3847/2515-5172/ac053c}

\bibitem[{Bavera {et~al.}(2020)Bavera, Fragos, Qin, Zapartas, Neijssel, Mandel,
  Batta, Gaebel, Kimball, \& Stevenson}]{Bavera:2019}
Bavera, S.~S., Fragos, T., Qin, Y., {et~al.} 2020, Astron. Astrophys., 635,
  A97, \dodoi{10.1051/0004-6361/201936204}

\bibitem[{Bavera {et~al.}(2021)}]{Bavera:2020uch}
Bavera, S.~S., {et~al.} 2021, Astron. Astrophys., 647, A153,
  \dodoi{10.1051/0004-6361/202039804}

\bibitem[{{Belczynski} {et~al.}(2010){Belczynski}, {Bulik}, {Fryer}, {Ruiter},
  {Valsecchi}, {Vink}, \& {Hurley}}]{Belczynski:2010ApJ}
{Belczynski}, K., {Bulik}, T., {Fryer}, C.~L., {et~al.} 2010, \apj, 714, 1217,
  \dodoi{10.1088/0004-637X/714/2/1217}

\bibitem[{Belczynski {et~al.}(2020)}]{Belczynski:2017gds}
Belczynski, K., {et~al.} 2020, Astron. Astrophys., 636, A104,
  \dodoi{10.1051/0004-6361/201936528}

\bibitem[{Belczynski {et~al.}(2022)Belczynski, Romagnolo, Olejak, Klencki,
  Chattopadhyay, Stevenson, Miller, Lasota, \& Crowther}]{Belczynski:2021zaz}
Belczynski, K., Romagnolo, A., Olejak, A., {et~al.} 2022, Astrophys. J., 925,
  69, \dodoi{10.3847/1538-4357/ac375a}

\bibitem[{{Biscoveanu} {et~al.}(2022){Biscoveanu}, {Callister}, {Haster}, {Ng},
  {Vitale}, \& {Farr}}]{Biscoveanu:2022}
{Biscoveanu}, S., {Callister}, T.~A., {Haster}, C.-J., {et~al.} 2022, arXiv
  e-prints, arXiv:2204.01578.
\newblock \doarXiv{2204.01578}

\bibitem[{Biscoveanu {et~al.}(2021)Biscoveanu, Isi, Vitale, \&
  Varma}]{Biscoveanu:2020are}
Biscoveanu, S., Isi, M., Vitale, S., \& Varma, V. 2021, Phys. Rev. Lett., 126,
  171103, \dodoi{10.1103/PhysRevLett.126.171103}

\bibitem[{Briel {et~al.}(2021)Briel, Eldridge, Stanway, Stevance, \&
  Chrimes}]{Briel:2021bpb}
Briel, M.~M., Eldridge, J.~J., Stanway, E.~R., Stevance, H.~F., \& Chrimes,
  A.~A. 2021.
\newblock \doarXiv{2111.08124}

\bibitem[{Broekgaarden(2021)}]{Broekgaarden:2021-zenodo-BHBH}
Broekgaarden, F.~S. 2021, {BHBH simulations from: Impact of Massive Binary Star
  and Cosmic Evolution on Gravitational Wave Observations II: Double Compact
  Object Mergers}, 1,  Zenodo, \dodoi{10.5281/zenodo.5651073}

\bibitem[{Broekgaarden {et~al.}(2019)Broekgaarden, Justham, de~Mink, Gair,
  Mandel, Stevenson, Barrett, Vigna-G\'omez, \&
  Neijssel}]{Broekgaarden:2019qnw}
Broekgaarden, F.~S., Justham, S., de~Mink, S.~E., {et~al.} 2019, Mon. Not. Roy.
  Astron. Soc., 490, 5228, \dodoi{10.1093/mnras/stz2558}

\bibitem[{{Broekgaarden} {et~al.}(2021){Broekgaarden}, {Berger}, {Neijssel},
  {Vigna-G{\'o}mez}, {Chattopadhyay}, {Stevenson}, {Chruslinska}, {Justham},
  {de Mink}, \& {Mandel}}]{Broekgaarden:2021iew}
{Broekgaarden}, F.~S., {Berger}, E., {Neijssel}, C.~J., {et~al.} 2021, \mnras,
  508, 5028, \dodoi{10.1093/mnras/stab2716}

\bibitem[{Broekgaarden {et~al.}(2022)}]{Broekgaarden:2021efa}
Broekgaarden, F.~S., {et~al.} 2022.
\newblock \doarXiv{2112.05763}

\bibitem[{{Callister} {et~al.}(2021){Callister}, {Haster}, {Ng}, {Vitale}, \&
  {Farr}}]{2021ApJ...922L...5C}
{Callister}, T.~A., {Haster}, C.-J., {Ng}, K. K.~Y., {Vitale}, S., \& {Farr},
  W.~M. 2021, \apjl, 922, L5, \dodoi{10.3847/2041-8213/ac2ccc}

\bibitem[{{Chattopadhyay} {et~al.}(2021){Chattopadhyay}, {Stevenson}, {Hurley},
  {Bailes}, \& {Broekgaarden}}]{Chattopadhyay:2020lff}
{Chattopadhyay}, D., {Stevenson}, S., {Hurley}, J.~R., {Bailes}, M., \&
  {Broekgaarden}, F. 2021, \mnras, 504, 3682, \dodoi{10.1093/mnras/stab973}

\bibitem[{Chia {et~al.}(2021)Chia, Olsen, Roulet, Dai, Venumadhav, Zackay, \&
  Zaldarriaga}]{Chia:2021mxq}
Chia, H.~S., Olsen, S., Roulet, J., {et~al.} 2021.
\newblock \doarXiv{2105.06486}

\bibitem[{{Claeys} {et~al.}(2014){Claeys}, {Pols}, {Izzard}, {Vink}, \&
  {Verbunt}}]{Claeys:2014}
{Claeys}, J.~S.~W., {Pols}, O.~R., {Izzard}, R.~G., {Vink}, J., \& {Verbunt},
  F.~W.~M. 2014, \aap, 563, A83, \dodoi{10.1051/0004-6361/201322714}

\bibitem[{Collette(2013)}]{collette_python_hdf5_2014}
Collette, A. 2013, Python and HDF5 (O'Reilly)

\bibitem[{{Deheuvels} {et~al.}(2014){Deheuvels}, {Do{\u{g}}an}, {Goupil},
  {Appourchaux}, {Benomar}, {Bruntt}, {Campante}, {Casagrande}, {Ceillier},
  {Davies}, {De Cat}, {Fu}, {Garc{\'\i}a}, {Lobel}, {Mosser}, {Reese},
  {Regulo}, {Schou}, {Stahn}, {Thygesen}, {Yang}, {Chaplin},
  {Christensen-Dalsgaard}, {Eggenberger}, {Gizon}, {Mathis},
  {Molenda-{\.Z}akowicz}, \& {Pinsonneault}}]{2014A&A...564A..27D}
{Deheuvels}, S., {Do{\u{g}}an}, G., {Goupil}, M.~J., {et~al.} 2014, \aap, 564,
  A27, \dodoi{10.1051/0004-6361/201322779}

\bibitem[{Di~Carlo {et~al.}(2020)}]{DiCarlo:2020lfa}
Di~Carlo, U.~N., {et~al.} 2020, Mon. Not. Roy. Astron. Soc., 498, 495,
  \dodoi{10.1093/mnras/staa2286}

\bibitem[{Dominik {et~al.}(2012)Dominik, Belczynski, Fryer, Holz, Berti, Bulik,
  Mandel, \& O'Shaughnessy}]{Dominik:2012kk}
Dominik, M., Belczynski, K., Fryer, C., {et~al.} 2012, Astrophys. J., 759, 52,
  \dodoi{10.1088/0004-637X/759/1/52}

\bibitem[{{Eldridge} {et~al.}(2017){Eldridge}, {Stanway}, {Xiao}, {McClelland},
  {Taylor}, {Ng}, {Greis}, \& {Bray}}]{BPASS:2017PASA}
{Eldridge}, J.~J., {Stanway}, E.~R., {Xiao}, L., {et~al.} 2017, \pasa, 34,
  e058, \dodoi{10.1017/pasa.2017.51}

\bibitem[{{Farmer} {et~al.}(2019){Farmer}, {Renzo}, {de Mink}, {Marchant}, \&
  {Justham}}]{2019ApJ...887...53F}
{Farmer}, R., {Renzo}, M., {de Mink}, S.~E., {Marchant}, P., \& {Justham}, S.
  2019, \apj, 887, 53, \dodoi{10.3847/1538-4357/ab518b}

\bibitem[{Farr {et~al.}(2017)Farr, Stevenson, Coleman~Miller, Mandel, Farr, \&
  Vecchio}]{Farr:2017uvj}
Farr, W.~M., Stevenson, S., Coleman~Miller, M., {et~al.} 2017, Nature, 548,
  426, \dodoi{10.1038/nature23453}

\bibitem[{{Fowler} \& {Hoyle}(1964)}]{1964ApJS....9..201F}
{Fowler}, W.~A., \& {Hoyle}, F. 1964, \apjs, 9, 201, \dodoi{10.1086/190103}

\bibitem[{{Fryer} {et~al.}(2012){Fryer}, {Belczynski}, {Wiktorowicz},
  {Dominik}, {Kalogera}, \& {Holz}}]{Fryer:2012ApJ}
{Fryer}, C.~L., {Belczynski}, K., {Wiktorowicz}, G., {et~al.} 2012, \apj, 749,
  91, \dodoi{10.1088/0004-637X/749/1/91}

\bibitem[{Fuller \& Ma(2019)}]{Fuller:2019sxi}
Fuller, J., \& Ma, L. 2019, Astrophys. J. Lett., 881, L1,
  \dodoi{10.3847/2041-8213/ab339b}

\bibitem[{{Fuller} {et~al.}(2019){Fuller}, {Piro}, \&
  {Jermyn}}]{Fuller:2019MNRAS}
{Fuller}, J., {Piro}, A.~L., \& {Jermyn}, A.~S. 2019, \mnras, 485, 3661,
  \dodoi{10.1093/mnras/stz514}

\bibitem[{Galaudage {et~al.}(2021)Galaudage, Talbot, Nagar, Jain, Thrane, \&
  Mandel}]{Galaudage:2021rkt}
Galaudage, S., Talbot, C., Nagar, T., {et~al.} 2021, Astrophys. J. Lett., 921,
  L15, \dodoi{10.3847/2041-8213/ac2f3c}

\bibitem[{{Gallegos-Garcia} {et~al.}(2021){Gallegos-Garcia}, {Berry},
  {Marchant}, \& {Kalogera}}]{Gallegos-Garcia:2021hti}
{Gallegos-Garcia}, M., {Berry}, C. P.~L., {Marchant}, P., \& {Kalogera}, V.
  2021, \apj, 922, 110, \dodoi{10.3847/1538-4357/ac2610}

\bibitem[{{Gehan} {et~al.}(2018){Gehan}, {Mosser}, {Michel}, {Samadi}, \&
  {Kallinger}}]{2018A&A...616A..24G}
{Gehan}, C., {Mosser}, B., {Michel}, E., {Samadi}, R., \& {Kallinger}, T. 2018,
  \aap, 616, A24, \dodoi{10.1051/0004-6361/201832822}

\bibitem[{Gerosa {et~al.}(2018)Gerosa, Berti, O'Shaughnessy, Belczynski,
  Kesden, Wysocki, \& Gladysz}]{Gerosa:2018wbw}
Gerosa, D., Berti, E., O'Shaughnessy, R., {et~al.} 2018, Phys. Rev. D, 98,
  084036, \dodoi{10.1103/PhysRevD.98.084036}

\bibitem[{Gerosa \& Fishbach(2021)}]{Gerosa:2021mno}
Gerosa, D., \& Fishbach, M. 2021, Nature Astron., 5, 8,
  \dodoi{10.1038/s41550-021-01398-w}

\bibitem[{Gerosa {et~al.}(2013)Gerosa, Kesden, Berti, O'Shaughnessy, \&
  Sperhake}]{Gerosa:2013laa}
Gerosa, D., Kesden, M., Berti, E., O'Shaughnessy, R., \& Sperhake, U. 2013,
  Phys. Rev. D, 87, 104028, \dodoi{10.1103/PhysRevD.87.104028}

\bibitem[{Gerosa {et~al.}(2020)Gerosa, Vitale, \& Berti}]{Gerosa:2020bjb}
Gerosa, D., Vitale, S., \& Berti, E. 2020, Phys. Rev. Lett., 125, 101103,
  \dodoi{10.1103/PhysRevLett.125.101103}

\bibitem[{Harris {et~al.}(2020)Harris, Millman, van~der Walt, Gommers,
  Virtanen, Cournapeau, Wieser, Taylor, Berg, Smith, Kern, Picus, Hoyer, van
  Kerkwijk, Brett, Haldane, Fernández~del Río, Wiebe, Peterson,
  Gérard-Marchant, Sheppard, Reddy, Weckesser, Abbasi, Gohlke, \&
  Oliphant}]{2020NumPy-Array}
Harris, C.~R., Millman, K.~J., van~der Walt, S.~J., {et~al.} 2020, Nature, 585,
  357–362, \dodoi{10.1038/s41586-020-2649-2}

\bibitem[{{Hotokezaka} \& {Piran}(2017)}]{Hotokezaka:2017ApJ}
{Hotokezaka}, K., \& {Piran}, T. 2017, \apj, 842, 111,
  \dodoi{10.3847/1538-4357/aa6f61}

\bibitem[{{Hu} {et~al.}(2022){Hu}, {Zhu}, {Qin}, {Zhang}, {Liang}, \&
  {Shao}}]{Hu:2022ubh}
{Hu}, R.-C., {Zhu}, J.-P., {Qin}, Y., {et~al.} 2022, \apj, 928, 163,
  \dodoi{10.3847/1538-4357/ac573f}

\bibitem[{{Hunter}(2007)}]{2007CSE.....9...90H}
{Hunter}, J.~D. 2007, Computing in Science and Engineering, 9, 90,
  \dodoi{10.1109/MCSE.2007.55}

\bibitem[{{Hurley} {et~al.}(2000){Hurley}, {Pols}, \&
  {Tout}}]{Hurley:2000MNRASSSE}
{Hurley}, J.~R., {Pols}, O.~R., \& {Tout}, C.~A. 2000, \mnras, 315, 543,
  \dodoi{10.1046/j.1365-8711.2000.03426.x}

\bibitem[{{Hurley} {et~al.}(2002){Hurley}, {Tout}, \&
  {Pols}}]{Hurley:2002MNRASBSE}
{Hurley}, J.~R., {Tout}, C.~A., \& {Pols}, O.~R. 2002, \mnras, 329, 897,
  \dodoi{10.1046/j.1365-8711.2002.05038.x}

\bibitem[{{Kaspi} {et~al.}(2000){Kaspi}, {Lyne}, {Manchester}, {Crawford},
  {Camilo}, {Bell}, {D'Amico}, {Stairs}, {McKay}, {Morris}, \&
  {Possenti}}]{2000ApJ...543..321K}
{Kaspi}, V.~M., {Lyne}, A.~G., {Manchester}, R.~N., {et~al.} 2000, \apj, 543,
  321, \dodoi{10.1086/317103}

\bibitem[{Kluyver {et~al.}(2016)Kluyver, Ragan-Kelley, P{\'e}rez, Granger,
  Bussonnier, Frederic, Kelley, Hamrick, Grout, Corlay,
  {et~al.}}]{kluyver2016jupyter}
Kluyver, T., Ragan-Kelley, B., P{\'e}rez, F., {et~al.} 2016, in ELPUB, 87--90

\bibitem[{{Kurtz} {et~al.}(2014){Kurtz}, {Saio}, {Takata}, {Shibahashi},
  {Murphy}, \& {Sekii}}]{2014MNRAS.444..102K}
{Kurtz}, D.~W., {Saio}, H., {Takata}, M., {et~al.} 2014, \mnras, 444, 102,
  \dodoi{10.1093/mnras/stu1329}

\bibitem[{Kushnir {et~al.}(2016)Kushnir, Zaldarriaga, Kollmeier, \&
  Waldman}]{Kushnir:2016zee}
Kushnir, D., Zaldarriaga, M., Kollmeier, J.~A., \& Waldman, R. 2016, Mon. Not.
  Roy. Astron. Soc., 462, 844, \dodoi{10.1093/mnras/stw1684}

\bibitem[{{Liu} \& {Lai}(2021)}]{Liu:2020gif}
{Liu}, B., \& {Lai}, D. 2021, \mnras, 502, 2049, \dodoi{10.1093/mnras/stab178}

\bibitem[{{Mandel} \& {Broekgaarden}(2021)}]{MandelBroekgaarden:2021}
{Mandel}, I., \& {Broekgaarden}, F.~S. 2021, arXiv e-prints, arXiv:2107.14239,
  \dodoi{https://doi.org/10.1007/s41114-021-00034-3}

\bibitem[{Mandel \& de~Mink(2016)}]{Mandel:2015qlu}
Mandel, I., \& de~Mink, S.~E. 2016, Mon. Not. Roy. Astron. Soc., 458, 2634,
  \dodoi{10.1093/mnras/stw379}

\bibitem[{Mandel \& Farmer(2022)}]{Mandel:2022hfr}
Mandel, I., \& Farmer, A. 2022, Phys. Rept., 955, 2201,
  \dodoi{10.1016/j.physrep.2022.01.003}

\bibitem[{Mandel \& Fragos(2020)}]{Mandel:2020lhv}
Mandel, I., \& Fragos, T. 2020, Astrophys. J. Lett., 895, L28,
  \dodoi{10.3847/2041-8213/ab8e41}

\bibitem[{{Mandel} \& {Smith}(2021)}]{Mandel:2021ewy}
{Mandel}, I., \& {Smith}, R. J.~E. 2021, \apjl, 922, L14,
  \dodoi{10.3847/2041-8213/ac35dd}

\bibitem[{Marchant {et~al.}(2016)Marchant, Langer, Podsiadlowski, Tauris, \&
  Moriya}]{Marchant:2016wow}
Marchant, P., Langer, N., Podsiadlowski, P., Tauris, T.~M., \& Moriya, T.~J.
  2016, Astron. Astrophys., 588, A50, \dodoi{10.1051/0004-6361/201628133}

\bibitem[{Mateu-Lucena {et~al.}(2021)Mateu-Lucena, Husa, Colleoni, Estell\'es,
  Garc\'\i{}a-Quir\'os, Keitel, Planas, \& Ramos-Buades}]{Mateu-Lucena:2021siq}
Mateu-Lucena, M., Husa, S., Colleoni, M., {et~al.} 2021.
\newblock \doarXiv{2105.05960}

\bibitem[{{McKernan} {et~al.}(2021){McKernan}, {Ford}, {Callister}, {Farr},
  {O'Shaughnessy}, {Smith}, {Thrane}, \& {Vajpeyi}}]{McKernan:2021correlation}
{McKernan}, B., {Ford}, K.~E.~S., {Callister}, T., {et~al.} 2021, arXiv
  e-prints, arXiv:2107.07551.
\newblock \doarXiv{2107.07551}

\bibitem[{{Neijssel} {et~al.}(2019){Neijssel}, {Vigna-G{\'o}mez}, {Stevenson},
  {Barrett}, {Gaebel}, {Broekgaarden}, {de Mink}, {Sz{\'e}csi}, {Vinciguerra},
  \& {Mandel}}]{Neijssel:2019}
{Neijssel}, C.~J., {Vigna-G{\'o}mez}, A., {Stevenson}, S., {et~al.} 2019,
  \mnras, 490, 3740, \dodoi{10.1093/mnras/stz2840}

\bibitem[{{Ng} {et~al.}(2018){Ng}, {Kruckow}, {Tauris}, {Lyne}, {Freire},
  {Ridolfi}, {Caiazzo}, {Heyl}, {Kramer}, {Cameron}, {Champion}, \&
  {Stappers}}]{2018MNRAS.476.4315N}
{Ng}, C., {Kruckow}, M.~U., {Tauris}, T.~M., {et~al.} 2018, \mnras, 476, 4315,
  \dodoi{10.1093/mnras/sty482}

\bibitem[{Nitz {et~al.}(2021)Nitz, Capano, Kumar, Wang, Kastha, Sch\"afer,
  Dhurkunde, \& Cabero}]{Nitz:2021uxj}
Nitz, A.~H., Capano, C.~D., Kumar, S., {et~al.} 2021, Astrophys. J., 922, 76,
  \dodoi{10.3847/1538-4357/ac1c03}

\bibitem[{Olejak \& Belczynski(2021)}]{Olejak:2021iux}
Olejak, A., \& Belczynski, K. 2021, Astrophys. J. Lett., 921, L2,
  \dodoi{10.3847/2041-8213/ac2f48}

\bibitem[{Olejak {et~al.}(2020)Olejak, Fishbach, Belczynski, Holz, Lasota,
  Miller, \& Bulik}]{Olejak:2020oel}
Olejak, A., Fishbach, M., Belczynski, K., {et~al.} 2020, Astrophys. J. Lett.,
  901, L39, \dodoi{10.3847/2041-8213/abb5b5}

\bibitem[{{Perez} \& {Granger}(2007)}]{2007CSE.....9c..21P}
{Perez}, F., \& {Granger}, B.~E. 2007, Computing in Science and Engineering, 9,
  21, \dodoi{10.1109/MCSE.2007.53}

\bibitem[{{Peters}(1964)}]{1964PhRv..136.1224P}
{Peters}, P.~C. 1964, Physical Review, 136, 1224,
  \dodoi{10.1103/PhysRev.136.B1224}

\bibitem[{{Portegies Zwart} \& {Verbunt}(1996)}]{1996A&A...309..179P}
{Portegies Zwart}, S.~F., \& {Verbunt}, F. 1996, \aap, 309, 179

\bibitem[{Qin {et~al.}(2018)Qin, Fragos, Meynet, Andrews, S\o{}rensen, \&
  Song}]{Qin:2018vaa}
Qin, Y., Fragos, T., Meynet, G., {et~al.} 2018, Astron. Astrophys., 616, A28,
  \dodoi{10.1051/0004-6361/201832839}

\bibitem[{{Qin} {et~al.}(2022){Qin}, {Wang}, {Wu}, {Meynet}, \&
  {Song}}]{Qin:2021vle}
{Qin}, Y., {Wang}, Y.-Z., {Wu}, D.-H., {Meynet}, G., \& {Song}, H. 2022, \apj,
  924, 129, \dodoi{10.3847/1538-4357/ac3982}

\bibitem[{Riley {et~al.}(2021)Riley, Mandel, Marchant, Butler, Nathaniel,
  Neijssel, Shortt, \& Vigna-Gomez}]{Riley:2020btf}
Riley, J., Mandel, I., Marchant, P., {et~al.} 2021, Mon. Not. Roy. Astron.
  Soc., 505, 663, \dodoi{10.1093/mnras/stab1291}

\bibitem[{Rodriguez {et~al.}(2016)Rodriguez, Zevin, Pankow, Kalogera, \&
  Rasio}]{Rodriguez:2016vmx}
Rodriguez, C.~L., Zevin, M., Pankow, C., Kalogera, V., \& Rasio, F.~A. 2016,
  Astrophys. J. Lett., 832, L2, \dodoi{10.3847/2041-8205/832/1/L2}

\bibitem[{Rodriguez {et~al.}(2020)}]{Rodriguez:2020viw}
Rodriguez, C.~L., {et~al.} 2020, Astrophys. J. Lett., 896, L10,
  \dodoi{10.3847/2041-8213/ab961d}

\bibitem[{Santoliquido {et~al.}(2020)Santoliquido, Mapelli, Bouffanais,
  Giacobbo, Di~Carlo, Rastello, Artale, \& Ballone}]{Santoliquido:2020bry}
Santoliquido, F., Mapelli, M., Bouffanais, Y., {et~al.} 2020, Astrophys. J.,
  898, 152, \dodoi{10.3847/1538-4357/ab9b78}

\bibitem[{{Schneider} {et~al.}(2015){Schneider}, {Izzard}, {Langer}, \& {de
  Mink}}]{Schneider:2015ApJ}
{Schneider}, F.~R.~N., {Izzard}, R.~G., {Langer}, N., \& {de Mink}, S.~E. 2015,
  \apj, 805, 20, \dodoi{10.1088/0004-637X/805/1/20}

\bibitem[{Shao \& Li(2022)}]{Shao:2022jzo}
Shao, Y., \& Li, X.-D. 2022.
\newblock \doarXiv{2203.14529}

\bibitem[{{Sipior} {et~al.}(2004){Sipior}, {Portegies Zwart}, \&
  {Nelemans}}]{2004MNRAS.354L..49S}
{Sipior}, M.~S., {Portegies Zwart}, S., \& {Nelemans}, G. 2004, \mnras, 354,
  L49, \dodoi{10.1111/j.1365-2966.2004.08373.x}

\bibitem[{Spruit(2002)}]{Spruit:2001tz}
Spruit, H.~C. 2002, Astron. Astrophys., 381, 923,
  \dodoi{10.1051/0004-6361:20011465}

\bibitem[{Stevenson {et~al.}(2017)Stevenson, Berry, \&
  Mandel}]{Stevenson:2017dlk}
Stevenson, S., Berry, C. P.~L., \& Mandel, I. 2017, Mon. Not. Roy. Astron.
  Soc., 471, 2801, \dodoi{10.1093/mnras/stx1764}

\bibitem[{Stevenson {et~al.}(2015)Stevenson, Ohme, \&
  Fairhurst}]{Stevenson:2015bqa}
Stevenson, S., Ohme, F., \& Fairhurst, S. 2015, Astrophys. J., 810, 58,
  \dodoi{10.1088/0004-637X/810/1/58}

\bibitem[{{Stevenson} {et~al.}(2019){Stevenson}, {Sampson}, {Powell},
  {Vigna-G{\'o}mez}, {Neijssel}, {Sz{\'e}csi}, \& {Mandel}}]{Stevenson:2019rcw}
{Stevenson}, S., {Sampson}, M., {Powell}, J., {et~al.} 2019, \apj, 882, 121,
  \dodoi{10.3847/1538-4357/ab3981}

\bibitem[{{Stevenson} {et~al.}(2017){Stevenson}, Vigna-G{\'o}mez, Mandel,
  Barrett, Neijssel, Perkins, \& de~Mink}]{Stevenson:2017tfq}
{Stevenson}, S., Vigna-G{\'o}mez, A., Mandel, I., {et~al.} 2017, Nature
  Communications, 8, 14906, \dodoi{10.1038/ncomms14906}

\bibitem[{Talbot \& Thrane(2017)}]{Talbot:2017yur}
Talbot, C., \& Thrane, E. 2017, Phys. Rev. D, 96, 023012,
  \dodoi{10.1103/PhysRevD.96.023012}

\bibitem[{Tang {et~al.}(2020)Tang, Eldridge, Stanway, \& Bray}]{Tang:2019qhn}
Tang, P.~N., Eldridge, J.~J., Stanway, E.~R., \& Bray, J.~C. 2020, Mon. Not.
  Roy. Astron. Soc., 493, L6, \dodoi{10.1093/mnrasl/slz183}

\bibitem[{{Tauris} \& {Sennels}(2000)}]{2000A&A...355..236T}
{Tauris}, T.~M., \& {Sennels}, T. 2000, \aap, 355, 236.
\newblock \doarXiv{astro-ph/9909149}

\bibitem[{{Team COMPAS: Riley} {et~al.}(2022){Team COMPAS: Riley}, {Agrawal},
  {Barrett}, {Boyett}, {Broekgaarden}, {Chattopadhyay}, {Gaebel}, {Gittins},
  {Hirai}, {Howitt}, {Justham}, {Khandelwal}, {Kummer}, {Lau}, {Mandel}, {de
  Mink}, {Neijssel}, {Riley}, {van Son}, {Stevenson}, {Vigna-Gomez},
  {Vinciguerra}, {Wagg}, \& {Willcox}}]{COMPAS:2021methodsPaper}
{Team COMPAS: Riley}, J., {Agrawal}, P., {Barrett}, J.~W., {et~al.} 2022,
  \apjs, 258, 34, \dodoi{10.3847/1538-4365/ac416c}

\bibitem[{{Toonen} {et~al.}(2018){Toonen}, {Perets}, {Igoshev}, {Michaely}, \&
  {Zenati}}]{2018A&A...619A..53T}
{Toonen}, S., {Perets}, H.~B., {Igoshev}, A.~P., {Michaely}, E., \& {Zenati},
  Y. 2018, \aap, 619, A53, \dodoi{10.1051/0004-6361/201833164}

\bibitem[{Vajpeyi {et~al.}(2022)Vajpeyi, Smith, \& Thrane}]{Vajpeyi:2022dvs}
Vajpeyi, A., Smith, R., \& Thrane, E. 2022.
\newblock \doarXiv{2203.13406}

\bibitem[{van~den Heuvel {et~al.}(2017)van~den Heuvel, Portegies~Zwart, \&
  de~Mink}]{vandenHeuvel:2017pwp}
van~den Heuvel, E. P.~J., Portegies~Zwart, S.~F., \& de~Mink, S.~E. 2017, Mon.
  Not. Roy. Astron. Soc., 471, 4256, \dodoi{10.1093/mnras/stx1430}

\bibitem[{van Son {et~al.}(2020)van Son, de~Mink, Broekgaarden, Renzo, Justham,
  Laplace, Moran-Fraile, Hendriks, \& Farmer}]{vanSon:2020zbk}
van Son, L. A.~C., de~Mink, S.~E., Broekgaarden, F.~S., {et~al.} 2020,
  Astrophys. J., 897, 100, \dodoi{10.3847/1538-4357/ab9809}

\bibitem[{van Son {et~al.}(2021)van Son, de~Mink, Callister, Justham, Renzo,
  Wagg, Broekgaarden, Kummer, Pakmor, \& Mandel}]{vanSon:2021zpk}
van Son, L. A.~C., de~Mink, S.~E., Callister, T., {et~al.} 2021.
\newblock \doarXiv{2110.01634}

\bibitem[{{\noopsort{Van Rossum}}{van Rossum}(1995)}]{CS-R9526}
{\noopsort{Van Rossum}}{van Rossum}, G. 1995, Python tutorial, Tech. Rep.
  CS-R9526, Centrum voor Wiskunde en Informatica (CWI), Amsterdam

\bibitem[{{Venkatraman Krishnan} {et~al.}(2020){Venkatraman Krishnan},
  {Bailes}, {van Straten}, {Wex}, {Freire}, {Keane}, {Tauris}, {Rosado},
  {Bhat}, {Flynn}, {Jameson}, \& {Os{\l}owski}}]{Venkatraman:2020}
{Venkatraman Krishnan}, V., {Bailes}, M., {van Straten}, W., {et~al.} 2020,
  Science, 367, 577, \dodoi{10.1126/science.aax7007}

\bibitem[{{Vigna-G{\'o}mez} {et~al.}(2018){Vigna-G{\'o}mez}, {Neijssel},
  {Stevenson}, {Barrett}, {Belczynski}, {Justham}, {de Mink}, {M{\"u}ller},
  {Podsiadlowski}, {Renzo}, {Sz{\'e}csi}, \& {Mandel}}]{Vigna-Gomez:2018dza}
{Vigna-G{\'o}mez}, A., {Neijssel}, C.~J., {Stevenson}, S., {et~al.} 2018,
  \mnras, 481, 4009, \dodoi{10.1093/mnras/sty2463}

\bibitem[{Virtanen {et~al.}(2020)Virtanen, Gommers, Oliphant, Haberland, Reddy,
  Cournapeau, Burovski, Peterson, Weckesser, Bright, {van der Walt}, Brett,
  Wilson, Millman, Mayorov, Nelson, Jones, Kern, Larson, Carey, Polat, Feng,
  Moore, {VanderPlas}, Laxalde, Perktold, Cimrman, Henriksen, Quintero, Harris,
  Archibald, Ribeiro, Pedregosa, {van Mulbregt}, \& {SciPy 1.0
  Contributors}}]{2020SciPy-NMeth}
Virtanen, P., Gommers, R., Oliphant, T.~E., {et~al.} 2020, Nature Methods, 17,
  261, \dodoi{10.1038/s41592-019-0686-2}

\bibitem[{Vitale {et~al.}(2017{\natexlab{a}})Vitale, Gerosa, Haster,
  Chatziioannou, \& Zimmerman}]{Vitale:2017cfs}
Vitale, S., Gerosa, D., Haster, C.-J., Chatziioannou, K., \& Zimmerman, A.
  2017{\natexlab{a}}, Phys. Rev. Lett., 119, 251103,
  \dodoi{10.1103/PhysRevLett.119.251103}

\bibitem[{Vitale {et~al.}(2017{\natexlab{b}})Vitale, Lynch, Sturani, \&
  Graff}]{Vitale:2015tea}
Vitale, S., Lynch, R., Sturani, R., \& Graff, P. 2017{\natexlab{b}}, Class.
  Quant. Grav., 34, 03LT01, \dodoi{10.1088/1361-6382/aa552e}

\bibitem[{Waskom \& the seaborn~development team(2020)}]{waskom2020seaborn}
Waskom, M., \& the seaborn~development team. 2020, mwaskom/seaborn, latest,
  Zenodo, \dodoi{10.5281/zenodo.592845}

\bibitem[{{W}es {M}c{K}inney(2010)}]{mckinney-proc-scipy-2010}
{W}es {M}c{K}inney. 2010, in {P}roceedings of the 9th {P}ython in {S}cience
  {C}onference, ed. {S}t\'efan van~der {W}alt \& {J}arrod {M}illman, 56 -- 61,
  \dodoi{10.25080/Majora-92bf1922-00a}

\bibitem[{Wysocki {et~al.}(2019)Wysocki, Lange, \&
  O'Shaughnessy}]{Wysocki:2018mpo}
Wysocki, D., Lange, J., \& O'Shaughnessy, R. 2019, Phys. Rev. D, 100, 043012,
  \dodoi{10.1103/PhysRevD.100.043012}

\bibitem[{{Zaldarriaga} {et~al.}(2018){Zaldarriaga}, {Kushnir}, \&
  {Kollmeier}}]{2018MNRAS.473.4174Z}
{Zaldarriaga}, M., {Kushnir}, D., \& {Kollmeier}, J.~A. 2018, \mnras, 473,
  4174, \dodoi{10.1093/mnras/stx2577}

\bibitem[{Zevin \& Bavera(2022)}]{Zevin:2022wrw}
Zevin, M., \& Bavera, S.~S. 2022.
\newblock \doarXiv{2203.02515}

\bibitem[{Zevin {et~al.}(2020)Zevin, Berry, Coughlin, Chatziioannou, \&
  Vitale}]{Zevin:2020gxf}
Zevin, M., Berry, C. P.~L., Coughlin, S., Chatziioannou, K., \& Vitale, S.
  2020, Astrophys. J. Lett., 899, L17, \dodoi{10.3847/2041-8213/aba8ef}

\end{thebibliography}

\section{Appendix}

\subsection{Zooming in on some of the model variations}
\label{app-fraction-MRR-outliers}
Three sets of stellar evolution models particularly stand out because of their mass-ratio reversal (\ac{MRR}) rates presented in Fig.~\ref{fig:FractionMRRbinaryBHs}.    
First, model B (which assumes a low mass transfer efficiency, $\beta$, arbitrarily fixed to $\beta = 0.25$) predicts one of the lowest \ac{MRR} fractions of $\pMRR^{\rm{det}} \approx 0.5$ and $\pMRR^{\rm{0}} \approx 0.4$ in Fig.~\ref{fig:FractionMRRbinaryBHs}.  
In these models the accreting star can at most accept $25\%$ of the mass from the donating star during a stable mass transfer phase. 
This mostly affects the first mass transfer phase as typically the second, reversed, mass transfer phase involves a compact object and is assumed to be Eddington limited.  
In our fiducial simulations the accretion efficiency instead is determined by the thermal timescale of the accretor \citep[][]{COMPAS:2021methodsPaper}, which, for binary black holes in our simulation, typically leads to almost fully conservative $\beta \approx 1$ mass transfer in the first mass transfer phase in the majority of our binary black hole progenitors (see Fig.~\ref{fig:DetailedEvolutionMRR}). 
In model B, the low $\beta$ significantly reduces the total amount that the initially less massive star accretes in our simulations. 
As a result the initially less massive star becomes less massive and the chance of \ac{MRR} is reduced.
Indeed, for increasing $\beta$ (models B, C, and D), we find an increasing fraction of \ac{MRR}.
The low mass transfer efficiency particularly lowers the stable mass transfer channel as this channel produces more of the near equal mass binary black hole mergers that can be \ac{MRR} in our fiducial model, but does not produce a massive enough secondary in the $\beta=0.25$ simulations. 
Instead, there is a visible rise from the `Other' channel, which is mostly coming from systems where the more massive initiates the first mass transfer phase on the main sequence, which is a slower mass transfer phase and therefore allows more overall mass accretion.  

Second, a subset of the models with strong Wolf-Rayet wind factors $f_{\rm{WR}}=5$ (model T) also have low \ac{MRR} fractions of $\pMRR^{\rm{det}} \approx 0.4$ and $\pMRR^{\rm{0}} \approx 0.11$, as shown in Fig.~\ref{fig:FractionMRRbinaryBHs}. 
The models with $f_{\rm{WR}}=5$ increase the mass loss through stellar winds during the Wolf-Rayet phase by a factor 5 compared to our fiducial model (A). 
As a result, the more massive star loses more mass during its Wolf-Rayet phase in model T and thus the star has a lower mass  when it finally undergoes a supernova. 
These lower mass stars are expected to produce larger amounts of ejecta during the supernova (or equivalently, have a lower fallback fraction) in the delayed \citet{Fryer:2012ApJ} supernova prescription implemented in all  COMPAS simulations presented here (except for model L). 
These lower mass stars also receive higher supernova natal kicks as the kicks are scaled down with the fallback fraction in the delayed prescription. 
This difference in kicks is particularly noticeable for the formation of \ac{MRR} binary black holes at higher (solar-like) metallicities as at lower metallicities the stars in both the fiducial model and the increased Wolf-Rayet factor model are typically massive enough ($\gtrsim 11\Msun$; see also \citealt{Fryer:2012ApJ,vanSon:2021zpk}) to receive (almost) complete fallback. 
For the higher metallicities case, on the other hand, the majority of binaries that in the fiducial model forms a binary black hole, disrupt during the first supernova as a result of the larger kicks in the models with large values of $f_{\rm{WR}}$.
This drastically reduces the number of \ac{MRR} systems at high metallicities as the \ac{MRR} systems typically have primaries with low pre-supernova masses that then receive these higher supernova natal kicks.  
As a result, particularly the \SFRD models that form  many stars with high average metallicities have much lower \ac{MRR} fractions. 
The same models also have significantly lower binary black hole merger rates, that fall below the currently observed rate as inferred from the third gravitational-wave catalog as shown earlier by \citet[][Fig.~2]{Broekgaarden:2021iew}.

Third, the models that assume the so-called `optimistic' common-envelope prescription (models F and K) stand out in Fig.~\ref{fig:FractionMRRbinaryBHs} as the classic common-envelope channel contributes significantly to the \ac{MRR} fraction. 
In this channel Hertzsprung Gap donor stars that engage in a common-envelope phase are allowed to survive (in the default `pessimistic case' such systems are assumed to undergo a merger).
This assumption significantly increases the total binary black hole mergers by adding systems that go through the common-envelope channel. 
A fraction of these systems undergo mass ratio reversal (in the first mass transfer), also increasing (decreasing) the relative fraction of \ac{MRR} from the common-envelope channel (only stable mass transfer) channel.

We provide for the interested reader example plots showing the impact on the mass evolution of several binary black hole progenitors for the majority model variations discussed in this section (models A, B, C, D, and T)  on our \href{https://github.com/FloorBroekgaarden/MRR_Project/tree/main/Figure1_individual_MRR_systems_detailed_evolution/detailedPlots/extraPlots}{GitHub}. 
We do not include the optimistic common-envelope models (F and K) as in these simulations binary black hole systems are added rather than that the evolution pathway is changed (see also \citealt{Broekgaarden:2021iew}).

\subsection{Mass Ratio Reversal BBH properties at birth}
\label{subsec:MRR-properties-birth}

\begin{figure*}
    \centering
    \includegraphics[width=1\textwidth]{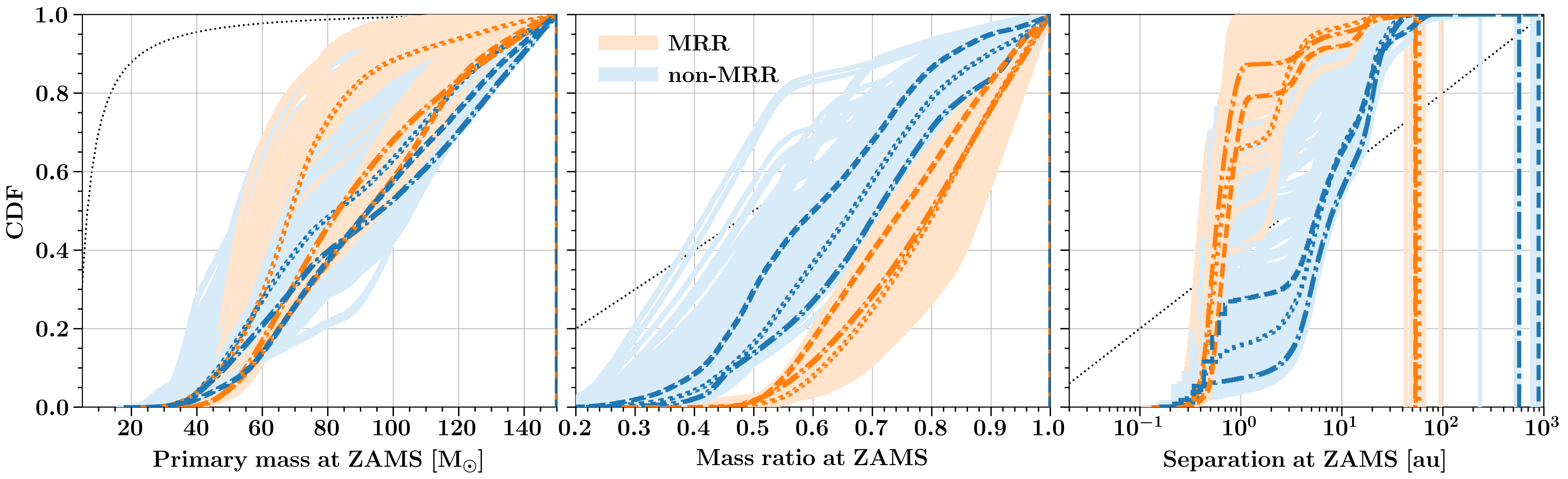}
    \caption{Cumulative distribution function of the zero-age main sequence more massive mass (left panel), mass ratio (center panel), and separation (right panel) for our 560 model realizations. 
    For each panel we show the distribution for mass ratio reversals (MRR; orange) and non mass ratio reversals (non-MRR; blue). 
    MRR binaries have more equal mass ratios and shorter separations at ZAMS compared to the overall BBH population. 
    In each panel for comparison in gray the birth distribution of all binaries in COMPAS are shown.
     Three realizations, `K123', `O312' and `T231' (see \href{https://github.com/FloorBroekgaarden/MRR_Project/blob/main/README.md}{Github}) are highlighted with a dotted, dash-dotted and dotted curve, respectively.
    \href{https://github.com/FloorBroekgaarden/MRR_Project/blob/main/Figure_Appendix_MRR_Properties_vs_nonMRR_ZAMS_CDFs_Panels/CDF_models_multiPanel_ZAMS.png}{\faFileImage} \href{https://github.com/FloorBroekgaarden/MRR_Project/blob/main/Figure_Appendix_MRR_Properties_vs_nonMRR_ZAMS_CDFs_Panels/FractionOfDetectable_BBHs.ipynb}{\faBook} }
    \label{fig:MRR-ZAMS-properties}
\end{figure*}

In Fig.~\ref{fig:MRR-ZAMS-properties} we show cumulative distributions of the initial properties of massive binaries that go on to form merging binary black holes.
In particular, we show the distributions for the more massive mass, the binary mass ratio and the orbital separation and split this up for \ac{MRR} and non-\ac{MRR} systems.
For each parameter, we additionally show the underlying distribution from which all massive binaries were sampled \citep[see][for further details]{Broekgaarden:2021efa}.
We highlight in Fig.~\ref{fig:MRR-ZAMS-properties} the initial parameters of the population of merging binary black holes that undergo \ac{MRR}.

We find that in all our models MRR binary black hole mergers form from binaries with birth properties that are similar to the total binary black hole (including both MRR and non MRR binaries) population. 
The most significant difference are that the MRR binary black hole population forms from a population of initial binaries with more equal mass binaries (such that they are closer to MRR initially) and smaller separations \citep[which favours more conservative mass transfer;][]{Schneider:2015ApJ}.

\end{document}